\documentclass[10pt]{article}
\usepackage{amsmath, a4wide, amssymb, mathrsfs, bbm, chicago}
\usepackage[english]{babel}
\usepackage[mathcal]{eucal}
\usepackage[totalwidth=500pt, totalheight=700pt]{geometry}

\numberwithin{equation}{section}

\newcommand{\real}[1]{\ensuremath{\mathbb{R}^{#1}}}
\newcommand{\complex}[1]{\ensuremath{\mathbb{C}^{#1}}}

\newcommand{\nninteger}{\ensuremath{\mathbb{N}}}
\newcommand{\Hilbert}{\ensuremath{\mathcal{H}}}

\newcommand{\vect}[2]{\ensuremath{{\mathbf #1}_{#2}}}

\newcommand{\Ldensity}[1]{\ensuremath{\mathscr{L}}}
\newcommand{\Hdensity}[1]{\ensuremath{\mathcal{H}}}

\newcommand{\vectr}[2]{\ensuremath{{\boldsymbol #1}_{#2}}}

\title{{\bfseries Physical Degrees of Freedom in Higgs Models}}
\author{M. Holman \\ Department of Physics and Astronomy, \hspace{0.05cm} Utrecht University, \\ Princetonplein 5,  \hspace{0.05cm} 3584 CC Utrecht, \hspace{0.05cm} The Netherlands \\ e-mail : {\ttfamily m.holman@phys.uu.nl}}

\begin{document}
\maketitle 
\begin{abstract}
\noindent Despite the clear-cut prediction and subsequent experimental detection of the weak interaction
bosons, the Higgs sector of the standard model of elementary particle physics has long remained one of its
most obscure features. Here, it is demonstrated through a very basic argument that standard accounts 
of the Higgs mechanism suffer from a serious conceptual consistency problem, in that they incorrectly identify 
physical degrees of freedom. The point at issue, is that the reasoning which leads to a removal of 
the unphysical excitation modes is valid in both phases of the theory - i.e. both after \emph{and before} 
the phase transition occurs. 
Consistently removing unphysical degrees of freedom implies a discrepancy in the number of physical degrees of 
freedom. In particular, the longitudinally polarized, massive gauge boson degrees of freedom do not have physical 
counterparts before the phase transition and are thus effectively ``created ex nihilio'' at the transition, 
within the context of ordinary Higgs models. Possible scenarios for removing the discrepancy are briefly considered.
The results obtained here strongly indicate that although standard, perturbative formulations of the Higgs mechanism
provide a convenient parametrization of electroweak physics over a certain range of scales, they cannot provide a 
sensible explanation of all relevant physical degrees of freedom involved. 
\end{abstract}

\section{Introduction}

\noindent Applications of notions of symmetry in various fields of human endeavour
can be traced back to ancient times, but it was the 20th century that has 
witnessed an ever more important role for symmetry principles in the formulation
of the laws of physics. In the course of these developments it has also been
recognized however that, in most cases, symmetries have only limited applicability.
For instance, to a very good approximation, the observable universe appears to be isotropic and
homogeneous on the largest distance scales, whereas this is manifestly not
the case on scales of (clusters of) galaxies and smaller. Similarly, when
viewed from sufficiently far, our own planet effectively corresponds to a
spherically symmetric distribution of matter, while according to popular belief, 
the possibility that earth has an even remotely near spherical symmetry was still 
contested in the time of Columbus - and for reasons not altogether unintelligible. 
At even smaller, atomic distance scales and at sufficiently low temperatures,
many elements display a regular lattice structure with translational and discrete rotational symmetries,
while no regular structure of this kind is usually discernible at larger distance scales.\\
In all of these examples, the (near) symmetry only manifests itself on a
particular (range of) scale(s), while it is absent on other scales.
As such, these types of symmetry seem less appropriate to investigate the amount
of symmetry structure in physical \emph{laws}\footnote{Modern quantum field theory
allows for the possibility that classical, ``macroscopic''
symmetries are \emph{anomalously broken} at the (arguably more fundamental) quantum
level, so that, roughly, it would appear that some symmetries could \emph{emerge}
from an underlying \emph{asymmetry} in ``microscopic'', dynamical physical laws. This aspect
of quantum field theory will be ignored in what follows however, basically 
because it crucially depends on the renormalization procedure for dealing
with divergent, i.e. ill-defined, quantities in the theory - a procedure which is the subject 
of a still ongoing debate concerning its logical-conceptual legitimacy. See e.g. \citeN{Feynman} or \citeN{CaoSch}.
More recently, Connes and collaborators have made attempts to obtain a better understanding of
renormalization, using advanced mathematical ideas. See \citeN{CoMa}.}. 
Concerning the latter, the recognition that many symmetries have only limited 
applicability may be represented in at least two ways. First, many symmetries 
are understood as only \emph{approximate} in present formulations of the laws of physics.
This is the case for instance for the Poincar\'e symmetries of Minkowski spacetime,
or the global $\mbox{SU}(2)$ symmetry of the internal isospin space associated with
the up- and down-quark. Second, symmetries may be exact with respect to the
dynamical laws of physics, but they may not be displayed by every \emph{solution}
to these dynamical laws. In this case one may also say that the symmetry is\enlargethispage*{5cm} 
exact at the dynamical level, but (partially) broken at the kinematical level. 
It may then happen that the parameters for a particular model of interest are non-dynamical and 
moreover always uniquely determine the solutions to the dynamical equations of the model, 
for a given set of initial conditions. In this case, solutions are either symmetric 
\newpage
\noindent or asymmetric, but no symmetry is broken dynamically (the 
inverse square force laws of classical gravity and electrostatics fall into this
category, for example). However, in the previous century it was found that there are
many situations of physical interest that can be described by models with
dynamically evolving parameters (not determined by the models themselves),
such that for certain ranges of parameter values,
\emph{all} stable solutions are asymmetric and are moreover not uniquely determined by these parameter values,
for a given set of initial conditions (in the theory of nonlinear differential equations, such a phenomenon
is referred to as the \emph{bifurcation} of solutions).
For such models, the symmetry (of solutions) can thus potentially be dynamically broken and
since the asymmetric solutions are not uniquely determined by parameter values,
one also speaks of \emph{spontaneous symmetry breaking} in this case.
In fact, although standard, the terminology ``symmetry breaking'' is somewhat
unfortunate in this context, even when it is clearly realized that no symmetry 
is broken at the dynamical level. This is because in some of the most important
applications of spontaneous symmetry breaking, the symmetries at issue are
\emph{local} symmetries and in that case no physical symmetry is broken whatsoever - at least classically. 
If there is any symmetry breaking at all, this occurs at most at the level of
\emph{description} of a particular classical physical system. It then seems more logical 
to refer to the defining event in question, at which a physical symmetry \emph{would}
be broken in the corresponding case in which the symmetry is global (see section \ref{SSBlocalU1}), simply as 
a ``phase transition'' and this practice will indeed be followed at times.
However, in order not to deviate too much from the extant literature on the 
point of semantics and also because it is necessary to reconsider the whole matter
in the light of quantum theory (where it in fact becomes rather obscure - essentially because
there is at present no fully satisfactory quantum description of physically relevant systems with local symmetries),
the standard terminology regarding spontaneous symmetry breaking 
will also be freely used (even in cases where this terminology is thus not really appropriate).\\
\noindent Many discussions of spontaneous symmetry breaking in the physics literature
start by considering the following potential for a complex scalar field $\phi$
\begin{equation}\label{ssbV1}
V(|\phi|^2) \; = \; m^2 |\phi|^2 \: + \: \frac{\lambda}{6}|\phi|^4 \qquad  \lambda > 0 \, , \; \; m^2 \in \real{}
\end{equation}
where $|\phi|^2 = \phi \phi^*$ (the only reason to consider a complex field in (\ref{ssbV1}), 
rather than two real fields, is mathematical convenience; also, the notation for the parameter 
$m^2$ is somewhat confusing but standard). For the time being, the dimension and the signature 
of the intrinsically flat space on which the scalar field is defined are left unspecified. 
The range of applications for the potential (\ref{ssbV1}) is wide. In the area of 
high-energy physics it basically describes the self-interactions
of the (still elusive) Higgs field, while in the area of condensed matter
physics it gives rise to the Landau-Ginzburg free energy, with
applications ranging from superconductivity and superfluidity to ferromagnetism.
The setup of this article is as follows. In section \ref{SSBglobalU1} the archetypal 
example of spontaneous symmetry breaking is reviewed, by considering what
happens to the ground state of the potential (\ref{ssbV1}), when some dynamical
process (the origins of which will be disregarded for the time being) changes
the parameter $m^2$ from a positive to a negative value. Here, the discussion
is in complete accordance with the standard physics literature. The analysis
is then repeated in section \ref{SSBlocalU1} for a certain ``minimally generalized''
model, in which the global phase invariance of the Lagrangian density associated
with (\ref{ssbV1}) has been replaced by a \emph{local} phase invariance.
At this point, the conclusions obtained deviate sharply from those in the
standard literature. In particular, it will be seen that the Higgs mechanism
for gauge boson mass generation in its usual formulation has a very serious conceptual flaw, in that
it effectively creates a \emph{new} physical degree of freedom. The discussion
is then further generalized to a non-Abelian setting in section \ref{extension},
where a similar result will be established. In all these cases, the flaw arises 
from applying the notion of gauge symmetry in different and inconsistent fashions 
before and after the occurrence of the phase transition.
As the treatment up until section \ref{extension} is essentially
classical, the main purpose of section \ref{qftcontext} is to point out that it
is unreasonable to expect that the physical inconsistency found for the standard formulation of the classical Higgs
mechanism does not persist upon quantization. In particular, it will be seen that
upon adopting one well known quantization method, the inconsistency carries over rigorously
to the quantum level. However, even if it were to turn out that the argument fails as a
quantum argument, it will also be seen in section \ref{qftcontext} that this would then require
- for conventional formulations of the Higgs mechanism - a rather strong faith in the applicability
of the principles of quantum theory (as well as the present-day standard interpretation of these
principles). Some non-standard formulations of the Higgs mechanism are considered in section \ref{altview},
especially in connection to their potential use for evading the main conclusion of sections \ref{SSBlocalU1}
and \ref{extension}. The main point here is that although it is indeed well imaginable that non-standard
formulations could be completely successful in this regard, they also lead to a number of further difficulties.
The results obtained here strongly indicate that so-called Higgs models, as usually understood, are
only able to provide a convenient parametrization of the electroweak sector of\enlargethispage*{10cm} 
the standard model and cannot, in fact, provide a sensible explanation of (all) 
the relevant physical degrees of freedom\footnote{The present work arose out of efforts to 
improve this author's understanding of the nature of gauge symmetries in general, the Higgs 
mechanism in particular and, to some extent, also of some earlier critical remarks on these
issues due to \citeANP{Earman1} \citeyear{Earman1,Earman3}.
After essentially completing it, he learned of two recent, similarly critical discussions 
of the Higgs mechanism due to \citeN{Smeenk} and \citeN{Lyre}. The similarities and, in particular, the
differences of these latter two treatments with the present work are discussed in section \ref{discussion}.}.

\newpage
\section{Spontaneously Broken Global $\mbox{U}(1)$-Symmetry}\label{SSBglobalU1}

\noindent It is a simple mathematical fact that for $m^2 > 0$ the potential (\ref{ssbV1})
has a single minimum at $\phi = 0$, in which the global U(1)-symmetry is unbroken\footnote{The potential
itself of course has a \emph{local} $\mbox{U}(1)$-symmetry as well, but upon also including kinetic terms 
for $\phi$, there is merely a global phase invariance, unless further modifications of the model are introduced.}, while 
for $m^2 < 0$, the minima are degenerate and form a circle of radius $|\phi| = v := \sqrt{- 3m^2 / \lambda}$.
If the potential (\ref{ssbV1}), for a continuous range of values for $m^2$ including
$0$, is taken to provide an approximate description of some physical system,
the symmetric ground state for $m^2 > 0$ will be ``triggered'' (by any small
perturbation of the system not included in the model) into one of the asymmetric
ground states as soon as $m^2$ becomes negative\footnote{This is certainly
true for any localized \emph{physical} system within the context of classical theory
(as no such system can be truely isolated). Within a cosmological context, the 
situation is slightly more subtle and here it could in fact be imagined that
the symmetry breaking is a truely spontaneous event.}.
Standard arguments may be applied to determine the physical spectrum of the model
defined by (\ref{ssbV1}). That is, $\phi$ is expressed as a small perturbation
around a minimum of $V$ and the nature of the physical degrees of freedom is
determined from the terms quadratic in the field in the subsequent expansion 
of the Lagrangian density, $\mathscr{L}$, given by
\begin{equation}\label{ssbL1}
\mathscr{L} \; = \; - |\partial \phi|^2 \: - \: V(|\phi|^2)
\end{equation}
where $|\partial \phi|^2 := (\partial_{\mu} \phi)(\partial^{\mu} \phi)^*$,
$\partial_{\mu} := \partial / \partial x^{\mu}$, $\mu = 0,1,2,3$, with indices
raised and lowered by the Minkowski metric (of $-+++$ signature)
and with a summation implied over repeated indices\footnote{Although interesting
applications of (\ref{ssbL1}) are known to exist also for lower dimensional
spacetimes, in view of the specific applications considered in the following
sections, attention is hereafter restricted to spacetime dimension four.}.
In view of the global phase symmetry of (\ref{ssbL1}), it may seem that
$\phi$ is most naturally expressed in terms of polar coordinates, but in fact
a canonical choice of coordinates for \emph{both} cases $m^2 > 0$ and $m^2 < 0$ does not exist.
Obviously, the physical spectrum cannot depend on a choice of coordinates
and one is free to simply \emph{choose} the coordinate system most apt for 
the case at hand. For $m^2 > 0$, the angular polar coordinate is ill-defined
at the single minimum of (\ref{ssbV1}) at $\phi = 0$, so that
the standard way to determine the physical spectrum runs into trouble upon
using polar coordinates in this case. In terms of Cartesian coordinates,
$(\phi_1 , \phi_2)$, a small perturbation, $\delta \phi := (\delta \phi_1 + i \delta \phi_2)/ \sqrt{2}$,
of the minimum at $\phi = 0$ obviously corresponds to two massive physical
degrees of freedom, $\delta \phi_1$, $\delta \phi_2$, of the same mass $m$ (here and in the following,
factors of $1 / \sqrt{2}$ appearing in expressions for perturbations are simply convenient normalization factors).\\
On the other hand, for $m^2 < 0$, a small perturbation of any ground state
in general leads to an unphysical spectrum upon using Cartesian coordinates,
unless the phase of the ground state is chosen to be exactly equal to an integer
multiple of $\pi / 2$ (i.e. corresponding to one of the two perturbations being
in the direction of the circle of degenerate minima). Again, this is simply
an artefact due to an inconvenient choice of coordinates. In terms of polar
coordinates, $(\rho, \theta)$, a small perturbation $\delta \phi := \delta \rho e^{i \delta \theta} / \sqrt{2}$
of a generic ground state $(\sqrt{2}v , \alpha)$, $\alpha \in \real{}$ is expressed as
\begin{equation}\label{perturb1}
\phi \; = \; \left( v + \delta \rho  / \sqrt{2} \right) e^{i (\alpha + \delta \theta)}
\end{equation}
and to second order in the perturbations $(\delta \rho , \delta \theta)$, the
Lagrangian density reads
\begin{equation}\label{2ndorderssbL1}
\mathscr{L}^{[2]} \; = \; - \frac{1}{2} (\partial \delta \rho)^2 \: - \: v^2 (\partial \delta \theta)^2 \: - \: \frac{1}{2} m_{\rho}^2 (\delta \rho)^2
\end{equation}
where $m_{\rho}^2 := - 2m^2 = 2 \lambda v^2 / 3$ and where an overall constant
term was discarded. Obviously, the radial perturbation, $\delta \rho$, represents
a massive degree of freedom with mass $m_{\rho}$, while the angular perturbation,
$\delta \theta$, corresponds to a massless degree of freedom. In fact, the
masslessness of the angular perturbation should be intuitively obvious,
since the potential is entirely phase-independent.
Another way to see this is to note that the set of degenerate minima of the
potential form a smooth, connected manifold, so that a small perturbation of
any minimum tangential to this manifold meets no resistance and therefore corresponds
to a massless mode.\\
The net effect of the symmetry breaking can thus be summarized by saying that
one physical degree of freedom becomes massless, while the other degree of
freedom, for a given value of $|m^2|$, becomes $\sqrt{2}$ times as heavy\footnote{From
a strictly logical perspective it remains somewhat unsettling what \emph{physical}
principle exactly determines the physicality of the radial and angular perturbations
in the broken symmetry phase. Obviously, polar coordinates are the \emph{natural}
coordinates to use in this case, given the symmetry properties of the potential.
But what principle determines their physicality ?\label{reshuffle}}.
As is well known, the appearance of massless excitations in the broken symmetry
phase also occurs in a quantum theory of fields, where it is known as the Goldstone 
phenomenon : given any (exact) continuous global symmetry group spontaneously broken into one 
of its subgroups, there is a massless spin-$0$ boson corresponding to each independent broken symmetry\footnote{\citeN{Goldstone}, \shortciteN{GoSaWe}. 
See also sections \ref{extension} and \ref{qftcontext}.}.
In fact, these features can be translated into rigorous statements within the
context of the Wightman approach to quantum field theory. Spontaneously broken
global symmetries are accompanied by vacuum degeneracy and are furthermore
not implemented by unitary transformations. They are thus not ``Wigner symmetries''
and this turns out to imply the presence of a massless one-quantum
state of helicity zero for each independent, spontaneously broken continuous symmetry (given the standard Wightman assumptions).
Although quite a few instances falling under the heading of Goldstone's theorem can be found in applications of
quantum field theory to condensed matter physics (e.g. spin waves in ferromagnets, phonons in crystals
and superfluids), the theorem poses a problem for attempts to break exact global 
symmetries within the context of high energy physics, as no massless elementary scalar particles 
are known to exist\footnote{In fact, the strong interaction displays an \emph{approximate} 
global chiral $\mbox{SU}(2) \times \mbox{SU}(2)$ symmetry, which is believed to 
be spontaneously broken into isospin $\mbox{SU}(2)$, thereby generating \emph{nearly}
massless (pseudo)-scalar particles, which are identified with the pions. It should be noted however that 
although the pions are the lightest known (pseudo)-scalar particles, they have a generic
mass that is still more than 250 times as large as the electron mass.\label{SSBQCD}}.
In what follows, the usual practice will be followed of referring to the massless
excitations in the broken symmetry phase - within the context of either classical
or quantum field theory - as ``Goldstone bosons (or modes)''.
So far, the discussion is essentially in accordance with standard treatments
of spontaneous symmetry breaking found in physics textbooks. The same is true
for the initial part of the discussion on spontaneously broken local symmetries
in the next section. Although reference could simply have been made to the
already existing literature in this regard, it was thought more useful to keep
the entire discussion somewhat more self-contained and to discuss matters from
scratch.

\section{Spontaneously Broken Local $\mbox{U}(1)$-Symmetry$\mbox{\large\protect\footnotemark}$ : the Abelian Higgs Model}\label{SSBlocalU1}

\addtocounter{footnote}{0}\footnotetext{As already explained in the introduction, the usual terminology regarding spontaneous symmetry breaking 
is mostly used in this article.}

\noindent As is well known, it is possible to generalize the Lagrangian density
(\ref{ssbL1}) in a certain ``minimal way'', so that it is invariant also under
\emph{local} phase rotations
\begin{equation}\label{gaugetrans1a}
\phi (x) \; \rightarrow \; e^{i \alpha (x)} \phi (x)
\end{equation}
of the complex scalar field, $\phi$, where $\alpha$ denotes some arbitrary
(differentiable) real-valued function defined on Minkowski spacetime.
This is achieved by introducing a ``gauge potential'', $A_{\mu}$, which is required
to transform as
\begin{equation}\label{gaugetrans1b}
A_{\mu} \; \rightarrow \; A_{\mu} \: + \: g^{-1} \partial_{\mu} \alpha
\end{equation}
in conjunction with the local phase rotation (\ref{gaugetrans1a}) of $\phi$,
and by replacing the ordinary derivative, $\partial_{\mu}$, in (\ref{ssbL1}),
by a covariant derivative, $D_{\mu}$, defined by
\begin{equation}
D_{\mu} \; := \; \partial_{\mu} \: - \: ig A_{\mu}
\end{equation}
In terms of physical interpretation, the real-valued ``coupling constant'',
$g$, governs the strength of interaction between the scalar field and the
newly introduced gauge potential. But in order for the gauge potential to
be associated with a physical, dynamical field, it is necessary to also introduce
kinetic terms - i.e. terms involving field-derivatives - for it. If the total Lagrangian
density is to remain invariant under the simultaneous transformations (\ref{gaugetrans1a}),
(\ref{gaugetrans1b}), any kinetic term for $A_{\mu}$ can enter this density
only in anti-symmetrized form\footnote{Although this seems quite obvious on intuitive
grounds, the statement in fact follows rigorously from an application of one
of Noether's theorems for local symmetries. See \citeANP{Utiyama1} \citeyear{Utiyama1,Utiyama2} for an explicit derivation.}
\begin{equation}\label{EMfieldstrength}
F_{\mu \nu} \; := \; \partial_{\mu} A_{\nu} \: - \: \partial_{\nu} A_{\mu}
\end{equation}
A minimal generalization of the Lagrangian density (\ref{ssbL1}), such that
this generalization is invariant under local phase rotations of the scalar
field, thus naturally suggests that the newly introduced field, $A_{\mu}$, is 
to be physically interpreted as an electromagnetic gauge potential and that
the total Lagrangian density should be
\begin{equation}\label{ssbL2}
\mathscr{L}_{\mbox{\scriptsize AHM}} \; = \; - \: |D \phi|^2 \: - \: V(|\phi|^2) \: - \: \frac{1}{4} F^2 
\end{equation}
Physically, this density represents a model - commonly referred to as the \emph{Abelian Higgs model} -
for a self-interacting, ``charged'' scalar field, $\phi$, coupled to an electromagnetic field, $F_{\mu \nu}$, 
by means of the gauge potential $A_{\mu}$. 
Before extending the analysis of the previous section to the present situation,
some further remarks concerning the steps that led to (\ref{ssbL2}) are in place.
First, no attempt will be made here to justify the ``gauge principle'', according
to which dynamical field equations \emph{should} be invariant under local symmetry
transformations\footnote{For some further discussion of this principle, see e.g. \citeN{Healey} or \citeN{Holman} 
and references therein. It should also be noted that, as is customary in many references nowadays, the terms
``gauge (symmetry) transformation'' and ``local symmetry transformation'' (either internal or external) are used
interchangeably in the present article. In particular, the phrase ``local gauge transformation'' is regarded as tautological.}.
However, the above reasoning shows that \emph{if} a Lagrangian density, which
possesses a certain global symmetry, is required to be invariant also under the
corresponding local symmetry (in which case the dynamical field equations will
certainly share this symmetry), this naturally leads to definite, nontrivial modifications
of this density. Second, no attempt will be made either to differentiate between the
subtle differences in interpretation that can be found in standard
references with regard to the notion of ``gauge symmetries''.
But what then is the common denominator in the standard interpretations of this notion ? 
In order to answer this question, the analysis of the previous
section is now repeated for the locally $\mbox{U}(1)$-invariant density
(\ref{ssbL2}). This will also lead to the main argument concerning physical
degrees of freedom in Higgs models.\\
For reasons that will become clear shortly, the case $m^2 < 0$ is considered
first this time. Just as in the corresponding globally $\mbox{U}(1)$-invariant
case, it is natural to use polar coordinates, $(\rho , \theta)$, when expanding
around a generic ground state $(\sqrt{2}v, \alpha)$ of the potential (\ref{ssbV1}).
To second order in the perturbations $(\delta \rho , \delta \theta)$ and up to an 
overall constant term, the Lagrangian density (\ref{ssbL2}) equals
\begin{eqnarray}
\mathscr{L}_{\mbox{\scriptsize AHM}}^{[2]} & = & - \frac{1}{2} (\partial \delta \rho)^2 \: - \: v^2 (\partial \delta \theta)^2 \: + \: 2g v^2 A^{\mu} \partial_{\mu} \delta \theta \: - \: \frac{1}{2} m_{\rho}^2 (\delta \rho)^2 \nonumber \\ 
                                           &   & - \frac{1}{2} m^2_A A^2 \: - \: \frac{1}{2} (\partial A)^2 \: + \: \frac{1}{2} (\partial_{\mu} A_{\nu}) (\partial^{\nu} A^{\mu})  \nonumber \\
                                           & = & \mathscr{L}^{[2]} \: + \: 2g v^2 A^{\mu} \partial_{\mu} \delta \theta \: + \: \mathscr{L}_{\mbox{\footnotesize Proca}} \label{2ndorderssbL2}
\end{eqnarray}
where $\mathscr{L}^{[2]}$ denotes the corresponding globally invariant second
order density (\ref{2ndorderssbL1}), where $\mathscr{L}_{\mbox{\footnotesize Proca}}$
stands for the familiar Proca density representative of a massive, classical 
spin-$1$ field (as implicitly defined by (\ref{2ndorderssbL2})) and where $m_A^2 := 2g^2v^2$.
At first sight, it thus appears that (\ref{2ndorderssbL2}) describes a massive 
mode, $\delta \rho$, and a Goldstone mode, $\delta \theta$, just as in the corresponding 
globally invariant case and that the net effect of the \emph{local} ``symmetry breaking'' 
is (i) the appearance of a term representing an interaction between the Goldstone 
mode and the gauge field - which itself arose as a consequence of the requirement 
of local phase symmetry - and (ii) the appearance of an additional degree of freedom 
for this gauge field, as a result of its becoming massive\footnote{It is recalled
that a vector field of nonzero mass (in ordinary spacetime) carries
three physical degrees of freedom : two physical states of ``transverse polarization'', 
also present for a massless vector field, and one physical state of ``longitudinal polarization''.\label{dofspin1field}}.\\
However, no use has yet been made of the local phase invariance of
(\ref{ssbL2}). In the corresponding globally invariant case, perturbing around
a generic ground state $(\sqrt{2}v , \alpha)$, results in an $\alpha$-independent
spectrum, precisely as a consequence of the global $\mbox{U}(1)$-symmetry.
In a similar fashion, the spectrum associated with (\ref{2ndorderssbL2}) \emph{should}
be locally phase independent precisely as a consequence of the local $\mbox{U}(1)$-symmetry.
Stated differently, just as one might choose any \emph{particular} value for
$\alpha$ and be sure that the resulting spectrum will not depend on this choice
in the globally invariant case, the local symmetry means that one might choose
any particular ``gauge'' for the perturbation \emph{of} a generic ground state and be
sure that the resulting spectrum will not depend on this gauge. In particular,
one may simply impose the so-called \emph{unitarity gauge} $\delta \theta = 0$ - or, equivalently, gauge
rotate the perturbation (\ref{perturb1}) according to $\phi (x) \rightarrow \exp (- i \delta \theta (x)) \phi (x) $ -
from which it is seen that, in the locally phase invariant case, the Goldstone
mode is simply a ``gauge artefact'' and disappears from the physical spectrum
altogether. The physical degrees of freedom associated with (\ref{2ndorderssbL2})
thus consist of one massive scalar amplitude mode of mass $m_{\rho}$ 
and three massive vector - i.e. gauge field - modes of mass $m_{A}$.
It is worth pointing out that this identification of the physical degrees of
freedom is in accord with standard treatments in the literature. As will
now be seen however, this is not so when it comes to identifying a possible
physical picture for the \emph{origins} of these degrees of freedom.\\
According to standard treatments, the appearance of the additional 
physical degree of freedom, corresponding to the longitudinal polarization of the gauge field,
should be attributed to the disappearance of the Goldstone mode. This is often
characterized by the catch-phrase that the gauge boson ``has eaten the Goldstone
boson and, as a result of this, has become massive''.
Such an account of things, generically referred to as the \emph{Higgs mechanism}, 
correctly describes the re-arrangement of physical degrees of
freedom in the process of going from a global invariance to the accompanying
local invariance. It is however not \emph{this} process that the standard treatments
pertain to - at least not explicitly. Indeed, for a given physical situation, 
what should be at most appropriate at the present classical level of discussion,
is a description in terms of \emph{either} a local invariance \emph{or} a global 
invariance that is not part of a local invariance.
Thus, for the specific locally $\mbox{U}(1)$-invariant model (\ref{ssbL2}), to
attribute the extra longitudinally polarized degree of freedom of the gauge
potential to a disappeared physical Goldstone mode is actually meaningless, since there
never was such a physical mode in the first place. Another way of saying this is that
the classical vacuum degeneracy, characteristic of the globally invariant model when 
$m^2 < 0$, is entirely absent in the corresponding locally invariant case
and there can thus be no Goldstone boson. This relation between the globally
$\mbox{U}(1)$-invariant model (\ref{ssbL1}) and the associated locally 
$\mbox{U}(1)$-invariant model (\ref{ssbL2}) generalizes to models with non-Abelian
symmetries. That is, there is no classical vacuum degeneracy for the gauged counterpart
of a globally invariant model with a degenerate set of minima (for a certain
range of parameter values of the model) and there are consequently no Goldstone
bosons. These remarks incidentally also clarify the statements made earlier, concerning 
the inappropriateness of the term ``symmetry breaking'' in the case where that symmetry 
is local and the treatment is classical\footnote{It is worth noting that these statements 
are in full agreement with standard discussions in the physics literature, in which, moreover, they are \emph{explicitly}
taken to be valid also as ``quantum descriptions'' (see e.g. \citeN{Englert}, section 3.1,
or \citeN{tHooft}, section 5.3). It should be noted however that it is possible to make a case that local symmetries 
\emph{can} be spontaneously broken in quantum theory (see e.g. \citeN{Strocchi}), although arguably not in any 
physical sense. See also section \ref{qftcontext}.\label{ssbquantum}}.\\
Returning now to the locally $\mbox{U}(1)$-invariant model (\ref{ssbL2}), with
its longitudinally polarized degree of freedom associated with the massive gauge boson when
$m^2 < 0$, it is clear that this degree of freedom should have a physical
counterpart when $m^2 > 0$ - at least, if the standard picture is adopted, according to
which a (cosmological) phase transition causes $m^2$ to change from a positive into a 
negative value. To determine the physical spectrum of (\ref{ssbL2})
for the unbroken symmetry phase, one again performs a small perturbation, $\delta \phi := (\delta \phi_1 + i \delta \phi_2)/ \sqrt{2}$,
of the minimum of the scalar field potential at $\phi = 0$ in terms of Cartesian coordinates.
As there is now no constant part in the modulus of the perturbed ground state (analogous
to the constant $v$ in (\ref{perturb1}) for the case $m^2 < 0$), the gauge field
picks up no mass this time. It thus initially seems that, in addition to the
two physical degrees of freedom of transversal polarization associated with
the massless gauge field, the physical spectrum consists of two massive, scalar
physical degrees of freedom of the same mass $m$, also present in the globally
invariant case. In fact, explicit statements to this effect can be found in
well known references on quantum field theory\footnote{See e.g. \citeN{ItZu},
section 12.5.3, \citeN{ChengLi}, section 8.3, or \citeN{Ryder}, section 8.3. 
Note that these discussions explicitly pertain to the \emph{locally} $\mbox{U}(1)$-invariant case.
See also section \ref{qftcontext}.\label{lit1}}. 
Again, no use has yet been made however of the \emph{local} phase invariance of the model (\ref{ssbL2}).
Indeed, it is intuitively clear that any two scalar perturbations of the same magnitude 
are merely different representations of the same physical perturbation, so that there is effectively
only one physical, scalar degree of freedom, i.e. a massive amplitude mode, of mass $m$. 
More rigorously, this follows from the fact that even though polar coordinates
are not actually defined \emph{at} the minimum $\phi = 0$ itself, there is nothing
that prevents expressing an arbitrary nontrivial perturbation \emph{of} this
minimum in polar coordinates (i.e. for the above perturbation in terms of
Cartesian coordinates, one simply puts $\delta \rho := \sqrt{\delta \phi_1^2 + \delta \phi_2^2}$,
$\delta \theta := \tan^{-1} ( \delta \phi_2 / \delta \phi_1 )$). Thus, just
as in the case $m^2 < 0$, the phase mode, $\delta \theta$, is nothing but a
gauge artefact, since it can be removed entirely by performing an appropriate gauge transformation.
These remarks, as well as the analogous remarks for the $m^2 < 0$ case,
just amount to the usual view on the meaning of gauge symmetries, already referred to earlier.
According to this view, field configurations related by gauge transformations, i.e. configurations lying in the
same ``gauge orbit'', should be physically identified and all that has been done in the foregoing, is to consistently
implement this standard view in both phases of the Abelian Higgs model.
One thus concludes that the physical degrees of freedom in the case
$m^2 > 0$ consist of \emph{one} massive scalar mode - just as in the broken symmetry phase - and two massless gauge
field modes. There is thus a discrepancy in the number of physical degrees
of freedom as compared for the two phases of the model. In particular,
the degree of freedom corresponding to the longitudinal polarization mode of 
the massive gauge boson in the broken symmetry phase does not have a physical counterpart 
in the unbroken symmetry phase, as things stand. It should be emphasized that these conclusions are in sharp 
contrast with standard accounts of the Higgs mechanism found in the physics
literature\footnote{In addition to the previously cited references, 
see e.g. \citeN{HaMa}, \citeN{Coleman} or \citeN{PeSch}.}.

\section{Some Further Considerations}

\noindent A query that can be raised about the foregoing argument is whether it is not ``overusing'' the gauge freedom.
The basic idea would be that once the gauge symmetry has been employed to get rid of a scalar field component, it can no longer be
used for the gauge field itself. As will be seen in this section, considerations of this nature do not affect the main 
conclusion of the present article, viz. the fact that there is a discrepancy in physical degrees of freedom for conventional
Higgs models, when these degrees of freedom are counted in the \emph{standard} manner and treated symmetrically with respect to the phase transition.
What these considerations do show however is that the order of different steps in the gauge fixing procedure is of crucial
importance for the outcome of the counting process.\\
Quite generically, there are two first class constraints for the specific Higgs model considered in the previous section, 
which, according to the usual treatments means that it is necessary to impose two gauge conditions to completely 
eliminate the gauge freedom. But it cannot simply be argued from this circumstance that there are $6 - 2 = 4$ physical degrees of freedom, 
because the two gauge conditions affect the scalar field and the gauge field simultaneously, as far as the full Lagrangian density 
is concerned. In terms of Cartesian coordinates, the scalar and gauge field degrees of freedom seem to have decoupled completely 
from the non-invariant, second order density associated with (\ref{ssbL2}) in the \emph{unbroken} symmetry phase, and one seems
to be left with two scalar field modes, $\delta \phi_1$ and $\delta \phi_2$, in addition to the four gauge field modes. The 
latter can be reduced to two independent modes by utilizing the gauge invariance of the electromagnetic part of the second order density, 
which is simply the original, pure Maxwell term. This is in agreement with the standard approaches. But, this way of reasoning\enlargethispage*{0.25cm} 
ignores the gauge invariance of the full density (\ref{ssbL2}). Concretely this means that in order to determine the physical 
spectrum in the usual way, one is at liberty to choose any gauge one wants for the scalar field perturbation, e.g. set $\delta \phi_2 = 0$ - 
just as this is typically done for the \emph{broken} symmetry phase (by necessity, it is argued, to get rid of the unwanted Goldstone 
mode). This also initially constrains the gauge field perturbation via (\ref{gaugetrans1b}). The key point however is that 
the Maxwell term in the perturbed, second order density is still \emph{fully gauge invariant} and it is only in this term that
the gauge field appears. Therefore, when it comes to counting independent degrees of freedom from the \emph{second order} density
- which is the standard quantity to determine such degrees of freedom from - one ends up this way with one physical scalar 
degree of freedom and two independent gauge field modes.\\
It may be useful here to make a comparison with the corresponding situation for the broken symmetry phase. Fixing the gauge so 
as to eliminate the Goldstone mode also constrains the gauge field perturbation via (\ref{gaugetrans1b}), although, usually, one just expresses 
the effect of the gauge transformation in this case by writing $A'$ instead of $A$ and then treat $A'$ as an \emph{unconstrained} 
variable, even though the Proca density is not gauge invariant. This then leads to three physical degrees of freedom (i.e. 
from four massive Klein-Gordon equations, constrained by a zero divergence condition), in addition to the physical scalar
Higgs mode. For the unbroken symmetry phase, the effect on $A$ of putting e.g. $\delta \phi_2 = 0$ is of course also simply that 
of a gauge transformation, $A \rightarrow A'$, as prescribed by (\ref{gaugetrans1b}). 
The field equations for $\delta \phi_1$ and $A'$ can then be determined from the perturbed density by neglecting higher order terms 
as usual. One ends up with one massive Klein-Gordon equation and the usual Maxwell equations in terms of $A'$, which are of 
course fully gauge invariant.\\
It might still be objected however that if a complete set of gauge conditions is imposed in order to implement a physical gauge 
on the potential \emph{first}, there is no longer any liberty to gauge rotate the scalar field (for instance, one could first 
impose the Lorentz condition $\partial \cdot A = 0$, and then use the remaining gauge freedom to set  $A_0$ equal to zero, 
thereby yielding two independent gauge field modes).
Although this is of course correct, it is not at all clear how one could end up with \emph{two} independent scalar degrees of 
freedom by proceeding this way. Initially, before imposing any gauge conditions, all scalar field configurations of the same 
magnitude lie in the same gauge orbit and are thus to be physically identified, according to the usual take on gauge symmetry.
So if it is argued that there are indeed two independent scalar degrees of freedom, the net effect on the scalar field of 
reducing the number of gauge field degrees of freedom to two, would be to \emph{increase} its number of physical degrees of freedom. 
The foregoing considerations show that the order of various steps in the gauge fixing is crucial for the outcome of the counting 
process. The main argument of the present work, i.e. that the steps which result in an identification of physical degrees of freedom 
in the broken symmetry phase also lead to an inconsistency when performed in a \emph{completely analogous} manner in the unbroken
symmetry phase, is unaffected by these considerations however.\\
At this point, the reader may still worry that something has gone wrong in the above 
analysis however. Even though the overall argument so far is logically straightforward 
and conceptually not very difficult, it was cast entirely within the language 
of classical field theory. Shouldn't these issues be properly addressed within 
the framework of \emph{quantum} field theory ? And if this is granted, wouldn't
that mean - especially when taking into consideration the intricate technicalities 
and conceptual difficulties that quantum field theory is notorious for - that
the above conclusions are at best premature ? In fact, not. As already pointed
out several times, the above discussion (if not the conclusions inferred from
it) is basically in agreement with many standard treatments of the Higgs mechanism
in the physics literature. In these treatments, classical field theory is (implicitly)
viewed as a lowest order - i.e. ``tree-level'' - description of \emph{quantum} field theory, as obtained 
by leaving out all terms in perturbation theory corresponding to
diagrams with closed loops. In particular, it is taken
for granted in these treatments that the standard way to infer the physical
particle spectrum of a quantum field theory - i.e. by inspecting the classical
Lagrangian density to second order in small field perturbations - is valid
and not affected by higher order ``quantum corrections''\footnote{\citeANP{tHooft}, \emph{ibid}, in his discussion of the Higgs mechanism,
for instance ignores (p. 35) ``contributions of loop diagrams, which represent the higher order quantum corrections to the field
equations''.}. Yet, it might still
be supposed that this is simply an omission of the standard treatments. In particular,
it might be wondered whether a non-standard account of vacuum degeneracy in a quantum
field theory with local symmetries could invalidate the above picture\footnote{Recall the comment in note \ref{ssbquantum}.
Also worth mentioning in this regard is the work of \citeN{ColeWein}, according to which vacuum degeneracy can arise in 
scalar electrodynamics, as result of including radiative corrections. See also \citeANP{ItZu}, \emph{ibid}, sct. 11-2-2.}. 
Moreover, at a much more basic level, it could also be wondered how to square the above findings with usual ideas in
quantum field theory, according to which the quantum theory of a single, complex scalar
field (considered just by itself) describes two types of quanta with opposite charge (in
fact, it is to these usual ideas that the standard references mentioned in the previous
notes appeal to - albeit mistakingly, within the present context). These worries
will be addressed in section \ref{qftcontext} however, where the question of 
whether the above conclusions are indeed likely to be affected within the context of
quantum field theory will be answered negatively. 
This thus means that an important question about the physical interpretation 
of the Higgs mechanism in its usual formulation, i.e. that concerning the physical origin of the extra 
degree of freedom associated with the massive gauge boson, is left unanswered.

\section{Extension to Non-Abelian Gauge Symmetries}\label{extension}

\noindent It is in principle straightforward to generalize the $\mbox{U}(1)$-invariant
models of the previous two sections to models with more general symmetry groups.
An obvious generalization of the potential (\ref{ssbV1}) - viewing this potential
now as pertaining to two real scalar fields, rather than a single complex field -
for $N$ real scalar fields, $\phi_1 , \cdots , \phi_N$, is 
\begin{equation}\label{ssbV2}
V(||\boldsymbol{\phi}||^2) \; = \; m^2 ||\boldsymbol{\phi}||^2 \: + \: \frac{\lambda}{6}||\boldsymbol{\phi}||^4 \qquad  \lambda > 0 \, , \; \; m^2 \in \real{}
\end{equation}
where vector notation was used and where $||\boldsymbol{\phi}||^2 := \sum_i \phi^2_i$.
As is easily verified and in complete analogy with the discussion in section
\ref{SSBglobalU1}, for $m^2 > 0$ the potential (\ref{ssbV2}) has a single minimum
at $\boldsymbol{\phi} = 0$, in which its $\mbox{O}(N)$-symmetry is unbroken,
while for $m^2 < 0$, the minima are degenerate and form a so-called \emph{vacuum manifold},
which for the present situation is a sphere, $S_v^{N-1}$, of radius $||\boldsymbol{\phi}|| = v := \sqrt{-3m^2/ \lambda}$. 
In the latter case, the generalization of the (commutative) rotational group
in two dimensions to a general rotational group in $N$ dimensions is seen to
lead to new effects for $N>2$, since selecting any one point, $\boldsymbol{\phi}_0 \in S_v^{N-1}$,
as the ground state leaves a residual $\mbox{O}(N-1)$-symmetry, corresponding
to rotations in the hyperplane perpendicular to $\boldsymbol{\phi}_0$.
As should be intuitively obvious, the remaining $N-1$ independent rotations,
in directions tangential to $S_v^{N-1}$, correspond to Goldstone modes. 
Another novelty brought about by the generalization of the potential (\ref{ssbV1})
to (\ref{ssbV2}) is that it becomes almost imperative to consider still further
generalizations. Indeed, in the above discussion it was tacitly assumed that
the $N$ scalar fields transform according to the defining, i.e. $N$-dimensional
vector, representation of $\mbox{O}(N)$. But, for a given value of $N$, it
may be possible to consider other, inequivalent representations of other
symmetry groups. For instance, for $N=6$, instead of transforming in the
defining representation of $\mbox{SO}(6)$ (taking the identity component
of the full invariance group, $\mbox{O}(6)$, for simplicity), the scalar fields
may be taken to transform in the adjoint representation of $\mbox{SO}(4)$.
Alternatively, and more in line with common practice, one may take a particular
symmetry group as somehow fundamental and consider the natural generalizations
of (\ref{ssbV2}) associated with all possible irreducible representations
of this group (in the case of $\mbox{U}(1)$ this does not lead to any novelties,
since the irreducible representations of this group are all one-dimensional).
For instance, instead of considering the real three-dimensional defining representation
of $\mbox{SO}(3)$ (again taking the identity component of $\mbox{O}(3)$ for simplicity), 
one could consider the complex two-dimensional spinor representation of this
group - i.e. the defining representation for its universal covering group,
$\mbox{SU}(2)$. In fact, in this case, it would seem natural to postulate
that the presumed underlying invariance is that of the group $\mbox{SU}(2)$, 
rather than $\mbox{SO}(3)$, in view of the simple connectivity of the former.
The number of possible ways in which the model (\ref{ssbV1}) can be naturally generalized
thus appears quite considerate.\\
Consider now the potential (\ref{ssbV2}) and suppose that the real scalar fields
$\phi_1 , \cdots , \phi_N$ transform according to an irreducible representation
of some special orthogonal group $\mbox{SO}(N')$. In other words, the potential
is supposed to be part of a theory with an underlying local $\mbox{SO}(N')$-invariance.
The restriction to the group $\mbox{SO}(N')$ is of no fundamental significance
here and is made only for the sake of simplicity. 
The aim is now to extend the discussion of section \ref{SSBlocalU1}.
To this end, let $t_a$ denote the generators of some particular $N$-dimensional, 
real irreducible representation of $\mbox{SO}(N')$, $[t_a , t_b] = - f_{ab}^{\mspace{16mu}c} t_c$, 
with $f_{abc}$ denoting the structure constants for $\mbox{SO}(N')$, $a, b, c = 1 , \cdots , N'(N'-1)/2$.
Then every local $\mbox{SO}(N')$-rotation of $\boldsymbol{\phi}$ can be expressed as\footnote{For compact,
connected Lie groups, such as $\mbox{SO}(N)$, $\mbox{U}(N)$ or $\mbox{SU}(N)$,
every group element can be written as an exponential of a Lie algebra element and
any matrix representative for it can similarly be expressed as the exponential
of a matrix representative of that Lie algebra element (by commutativity of the
diagram that relates the exponential map to any group homomorphism).}
\begin{equation}\label{gaugetrans2a}
\boldsymbol{\phi} (x) \; \rightarrow \; \Omega (x) \boldsymbol{\phi} (x) \; := \; e^{\alpha^a (x) t_a} \boldsymbol{\phi} (x)
\end{equation}
where the $N'(N'-1)/2$ functions $\alpha^a$ are real-valued and differentiable,
but further arbitrary, and where the Einstein summation convention has been
employed for contracted indices. In complete analogy with the discussion in
section \ref{SSBlocalU1}, the Lagrangian density associated with (\ref{ssbV2})
can be generalized in a certain ``minimal way'', such that it is invariant under
the local $\mbox{SO}(N')$-rotations (\ref{gaugetrans2a}) (rather than just
the subset of these rotations for which the $\alpha^a$ are constants).
This is again achieved by introducing a (Lie algebra valued) gauge potential,
$A_{\mu} := A_{\mu}^a t_a$, which is required to transform according to
\begin{equation}\label{gaugetrans2b}
A_{\mu} \; \rightarrow \; \Omega A_{\mu} \Omega^{-1} \: + \: g^{-1} ( \partial_{\mu} \Omega ) \Omega^{-1}
\end{equation}
in conjunction with the local rotation (\ref{gaugetrans2a}) of $\boldsymbol{\phi}$,
and by replacing the ordinary derivative, $\partial_{\mu}$, in the Lagrangian
density associated with (\ref{ssbV2}) by a covariant derivative, $D_{\mu}$, defined by
\begin{equation}\label{covder}
D_{\mu} \; := \; \partial_{\mu} \: - \: g A_{\mu}
\end{equation}
where $g \in \real{}$ again denotes a coupling constant. In order for $A_{\mu}$ 
to be associated with a physical, dynamical field, it is necessary to also add 
kinetic terms for it to the Lagrangian density and the gauge symmetry implies, 
once more, that any kinetic term for $A_{\mu}$ can enter this density only in 
anti-symmetrized form. However, this time there is a further restriction on the manner in which
derivative terms for $A_{\mu}$ can enter the Lagrangian density, due to the
non-Abelian character of the gauge symmetry, and, in fact, such derivative
terms can enter only via the field strength\footnote{Cf. \citeN{Utiyama1}.}
\begin{equation}
F_{\mu \nu} \; := \; - g^{-1} [D_{\mu} , D_{\nu}]
\end{equation}
or, in terms of components
\begin{equation}
F^a_{\mu \nu} \; := \; \partial_{\mu} A^a_{\nu} \: - \: \partial_{\nu} A^a_{\mu} \: + \: g f^a_{\mspace{12mu}bc} A^b_{\mu} A^c_{\nu}
\end{equation}
In analogy with (\ref{ssbL2}), a natural, minimal generalization of the density
associated with (\ref{ssbV2}), invariant under local $\mbox{SO}(N')$ rotations,
is then given by
\begin{equation}\label{ssbL3}
\mathscr{L}_{\mbox{\scriptsize NHM}} \; = \; - \: ||D \boldsymbol{\phi}||^2 \: - \: V(||\boldsymbol{\phi}||^2) \: - \: \frac{1}{2} \mbox{Tr} \, F^2 
\end{equation}
where the normalization condition $\mbox{Tr} \, t_a t_b = \delta_{ab}/2$ has been 
imposed on the generators. Incidentally, in the particular case where $f_{abc} = \epsilon_{abc}$,
the last term in (\ref{ssbL3}) is precisely the density proposed by \citeN{YaMi} 
- also on the basis of the analogy with electromagnetism - in their pioneering attempt
to deal with the $\mbox{SU}(2)$ isotopic spin invariance of the nuclear interactions. 
It is customary to refer to the model defined by the last term in (\ref{ssbL3}) 
also as a (pure) Yang-Mills model, and to the model defined by the total density 
as a \emph{non-Abelian Higgs model}.\\
From a historical perspective, a serious difficulty facing initial attempts
to apply Yang-Mills models to describe the nuclear interactions, was the 
short-ranged nature of these interactions. It seemed that the gauge potentials
in Yang-Mills models always had to be long-range, corresponding to massless quanta 
in the appropriate quantum field theory versions of these models (analogous 
to the photon in the theory of quantum electrodynamics). Although it was
known to be possible to include extra terms for the gauge potentials, thereby
causing them to effectively become short-range and to be associated with massive
quanta in the appropriate quantum field theory context, it was also known
that such extra terms would spoil the gauge invariance of Yang-Mills models, and
thereby, their renormalizability. In a somewhat similar fashion, and as already
mentioned, a serious difficulty facing initial attempts to apply potentials of the general symmetry
breaking type (\ref{ssbV2}) to describe high-energy phenomena, was the implied
existence of massless scalar quanta - i.e. the Goldstone bosons - associated
with the global symmetries of these potentials. The fact that combining
the two types of models into a model of the generic non-Abelian Higgs type (\ref{ssbL3})
makes it possible (see below) to simultaneously (i) incorporate massive Yang-Mills potentials without violating
gauge invariance and (ii) avoid physical massless scalar quanta, explains
the great theoretical appeal of these types of models.\\
To determine the physical spectrum for (\ref{ssbL3}) in the case
$m^2 < 0$, $\boldsymbol{\phi}$ is again expressed as a small perturbation
of a generic ground state, $\boldsymbol{\phi}_0$, of the potential (\ref{ssbV2})
\begin{equation}\label{perturb2}
\boldsymbol{\phi} \; = \; \boldsymbol{\phi}_0 \: + \: \delta \boldsymbol{\phi} / \sqrt{2}  \qquad \quad || \boldsymbol{\phi}_0 ||^2 \: = \: v^2
\end{equation}
and to second order in the fields and up to an overall constant, the Lagrangian
density (\ref{ssbL3}) then equals
\begin{eqnarray}\label{2ndorderssbL3}
\mathscr{L}^{[2]}_{\mbox{\scriptsize NHM}} & = & -  \frac{1}{2} (\partial \delta \boldsymbol{\phi})^2 \: + \: \sqrt{2} g A_{\mu}^a (t_a \boldsymbol{\phi}_0) \cdot (\partial^{\mu} \delta \boldsymbol{\phi}) 
                                                 \: - \: \frac{\lambda}{3} (\boldsymbol{\phi}_0 \cdot \delta \boldsymbol{\phi} )^2  \nonumber \\
                                           &   & - \: \frac{1}{2} m^2_{ab} A_{\mu}^a A^{b \mu} \: - \: \frac{1}{2} (\partial_{\mu} A_{\nu}^a)(\partial^{\mu} A^{\nu}_a)
                                                 \: + \: \frac{1}{2} (\partial_{\mu} A_{\nu}^a)(\partial^{\nu} A^{\mu}_a)
\end{eqnarray}
with $m^2_{ab} := g^2 (t_a \boldsymbol{\phi}_0) \cdot (t_b \boldsymbol{\phi}_0)$.
The matrix $m^2_{ab}$, being real and symmetric, is diagonalizable and by
forming appropriate linear combinations of the gauge potential components
with respect to the internal index (if necessary) a term of the exact same
form as the fourth term in (\ref{2ndorderssbL3}) can be obtained with $m^2_{ab}$
diagonal and with its nonzero eigenvalues equal to $g^2 v^2$.
This term is then seen to represent a collection of mass terms for the
gauge potentials, $A_{\mu}^a$, in exact correspondence to the $N''$ generators
$t_a$ that do not leave $\boldsymbol{\phi}_0$ invariant, i.e. those $t_a$
for which $t_a \boldsymbol{\phi}_0 \neq 0$, $0 < N'' \leq N'(N' - 1)/2$. The 
remaining $N'(N'-1)/2 - N''$ generators (corresponding to rotations in the hyperplane to which $\boldsymbol{\phi}_0$
is orthogonal) leave invariant $\boldsymbol{\phi}_0$ and the components of
$A_{\mu}$ in these directions of the Lie algebra remain massless. It is furthermore
seen from the second term that the scalar field gradients couple only to the
massive components of the gauge potential. Again, in complete analogy with the 
Abelian case, these massive gauge field components come in the place of the
scalar Goldstone modes of the corresponding globally invariant model. Put
differently, for each massive component, $A_{\mu}^a$, of the gauge potential,
the inner product $(t_a \boldsymbol{\phi}_0) \cdot (\partial_{\mu} \delta \boldsymbol{\phi})$
picks out the component of the scalar field perturbation, $\delta \boldsymbol{\phi}$,
in the direction of $t_a \boldsymbol{\phi}_0$, tangential to the vacuum manifold
$S^{N-1}_v$, and connecting $\boldsymbol{\phi}_0$ to an infinitesimally
rotated ground state. The perturbations tangential to $S^{N-1}_v$ represent
the $N-1$ Goldstone modes in the corresponding globally invariant model, but
in the present locally invariant context they are nothing but gauge artefacts,
which can be transformed away entirely by applying a suitable local symmetry
transformation (\ref{gaugetrans2a}). There are then essentially three main
possibilities, according to whether the dimension of the vacuum manifold,
$N-1$, is smaller than, equal to, or greater than $N'(N'-1)/2$, i.e. the dimension
of the symmetry group in the present context. In the first two cases, $N''=N-1$,
while in the third case $N''=N'(N'-1)/2$. In the first case, $N'(N'-1)/2 - (N-1)$
gauge fields thus remain massless, whereas all gauge fields acquire a mass in 
the last two cases. What distinguishes these last two cases is that in the 
third case there are more independent directions in $S^{N-1}_v$ than generators
of rotations and there are $N-1 - N'(N'-1)/2$ independent scalar field perturbations
orthogonal to any non-invariant ground state $\boldsymbol{\phi}_0$ and not in any 
symmetry direction. These perturbations correspond to gauge artefacts, but
they are not associated with any gauge boson. For instance, for $N'=3$, taking 
$N=3$ (real three-dimensional vector representation) leaves a residual $\mbox{U}(1)$-invariance with one corresponding massless
gauge boson, while taking $N=4$ (complex two-dimensional spin-$1/2$ representation)
or $N=6$ (complex three-dimensional spin-$1$ representation) leaves no unbroken 
symmetry and all three gauge fields acquire masses. In addition, in the last
case there are two independent scalar field perturbations which are gauge artefacts, 
but which are not in any symmetry direction. The physical spectrum for (\ref{ssbL3}) 
in the case $m^2 < 0$ thus consists of (i) one scalar excitation mode of mass $m_{\rho} = \sqrt{-2m^2}$, corresponding
to a perturbation, $\delta \boldsymbol{\phi}$, in the radial direction,
(ii) $N'(N'-1) - 2N''$ massless gauge field modes and (iii) $3N''$ massive
gauge field modes of the same mass $g^2 v^2$. It is worth noting that, apart
from the case in which the dimension of the vacuum manifold is larger than
the number of independent symmetry directions, these conclusions are in 
accordance with standard treatments in the physics literature.\\
If $m^2 > 0$, there is no vacuum degeneracy and the intuitive picture for
the case $m^2 < 0$, with only the radial scalar perturbation corresponding
to a physical excitation mode (i.e. the perturbation parallel to some chosen
vacuum), breaks down, as there are now infinitely many radial perturbations
for the unique ground state at zero field value. Again, this is simply a 
consequence of the fact that the natural coordinates for this case in view of
the symmetry properties of the potential, i.e. spherical coordinates, are not
defined at the origin of field space. The scalar field potential depends on 
a single variable only and there can thus be at most one massive scalar excitation
mode. This is the modulus (or amplitude) mode, $|| \delta \boldsymbol{\phi} ||$. The other
scalar excitation modes correspond to angular modes and appear only in the form
of kinetic terms in the Lagrangian density. These components of the perturbation can be
removed completely by performing a gauge transformation and are thus unphysical.
The physical spectrum for (\ref{ssbL3}) in the case $m^2 > 0$ thus consists
of one massive scalar excitation mode and $N'(N'-1)$ massless gauge field modes.
This is again in sharp contrast with most treatments in the physics literature.
What is more, there are now $N''$ physical degrees of freedom unaccounted for.\\
In physical applications, usually only the possibility $N-1 \leq N'(N'-1)/2$
is considered, so that the number of massive gauge fields equals $N-1$. In the
case of a potential with a generic $G$-invariance, the vacuum manifold then
equals the coset space, $\rho (G) / \rho (H)$, where $\rho : G \rightarrow \mbox{Gl}(V,\real{})$
denotes an irreducible representation of $G$ on the $N$-dimensional real vector 
space $V$ and the subgroup $H \leq G$ represents the residual symmetry group 
(that leaves invariant any chosen minimum)\footnote{It should be noted that it
is now possible to have more than one ``Higgs mode'' (cf. also \citeN{Huang}, section 4.6).
That is, if $N'' = \mbox{dim}(G) - \mbox{dim}(H)$ again denotes the number
of generators that do not leave invariant any chosen ground state in this
representation of $G$, there are a priori $N - N''$ ``Higgs modes'', in addition
to the $N''$ Goldstone modes. In the particular case of the potential (\ref{ssbV2})
however, with $G = \mbox{SO}(N')$ (as is basically appropriate for the electroweak sector of the standard 
model, since $\mbox{SU}(2) \simeq \mbox{SO}(3)$, $\mbox{U}(1) \simeq \mbox{SO}(2)$,
and a general irreducible representation of a direct product, $G_1 \times G_2$,
of two compact Lie groups, $G_1$, $G_2$, takes the form of a tensor product of
individual irreducible representations), one has $N'' = N - 1$ if $N - 1 \leq N'(N'-1)/2$ 
and $N''= N'(N'-1)/2$ if $N - 1 > N'(N'-1)/2$. But in the latter case, there is still only one
Higgs mode, as there are now $(N-1) - N'(N'-1)/2$ additional scalar field
directions corresponding to gauge artefacts but not associated with any
$\mbox{SO}(N')$ generator.}.
In fact, a particular instance of this general fact was already encountered
in the above discussion in which the defining representation of $\mbox{O}(N)$
was considered, since $S_v^{N-1} \simeq \mbox{O}(N) / \mbox{O}(N-1)$ (as can easily be verified).

\section{Quantum (Field) Theoretical Considerations}\label{qftcontext}

\noindent Upon consulting any thorough discussion on local symmetries and constraints\footnote{See e.g. \citeN{HenTei}.},
it would seem that it is in principle very straightforward to quantize field theories with local symmetries. All one really needs
to do is to extract the correct reduced phase space, $\Gamma_{\mbox{\scriptsize red}}$, by factoring out all the gauge orbits.
All observables of the theory, being gauge invariant, should be representable as functions on $\Gamma_{\mbox{\scriptsize red}}$
and since the so-called Dirac bracket equips $\Gamma_{\mbox{\scriptsize red}}$ with a symplectic structure,
standard canonical quantization techniques should in principle (fail to) be applicable in exactly the same fashion as for
field theories without gauge symmetries\footnote{See e.g. \citeN{Wald3} or \citeN{Giulini}.}.
However, it turns out to be very difficult to explicitly construct $\Gamma_{\mbox{\scriptsize red}}$ in most cases. Locally,
$\Gamma_{\mbox{\scriptsize red}}$ is determined by specifying a complete set of gauge conditions, as many as there are first
class constraints, $\gamma_a$, $a=1, \cdots, K'$, for some $K' \in \nninteger$.
But no complete set of global gauge conditions in general exists\footnote{Cf. \citeN{Gribov}, \citeN{Singer}.} and important global 
information may therefore be missed by proceeding this way. In a field theoretical setting, this essentially restricts the entire treatment to perturbation
theory in most cases. However, even if this difficulty is ignored and a route to a ``local'' quantization of $\Gamma_{\mbox{\scriptsize red}}$
is taken by specifying a complete set of gauge conditions on the original constraint surface, $\Gamma_{\mbox{\scriptsize c}}$, one runs into 
the problem that in many cases, ``physical'' gauges - such as the Coulomb gauge condition, $\boldsymbol{\nabla} \cdot \vect{A}{} = 0$, $A^0 = 0$,
in electrodynamics, which effectively singles out the degrees of freedom of transverse polarization - suffer from undesirable
features, such as the loss of manifest relativistic covariance or the loss of (weak) local commutativity - properties usually
deemed crucial for any Poincar\'e covariant quantum theory fundamentally based on fields\footnote{The Coulomb gauge actually
suffers from both mentioned difficulties. In the absence of sources,
given any gauge potential, $A_{\mu}$, in Maxwell theory, it is always possible, by making a suitable gauge transformation
(\ref{gaugetrans1b}), if neccesary, to impose the Coulomb gauge (i.e. first impose the Lorentz gauge $\partial \cdot A = 0$, 
by simply choosing $\alpha$ in (\ref{gaugetrans1b}) to be a solution to $\Box \alpha = - g (\partial \cdot A)$ and then perform 
a further gauge transformation, to set $A^0 = 0$; see e.g. \citeN{Wald2} for a discussion on how to achieve this in practice).
In the case of electrodynamics, this moreover completely fixes the gauge globally, thereby allowing for the ``global'' quantization
of $\Gamma_{\mbox{\scriptsize red}}$. In the non-Abelian case, Gribov obstructions arise and the Coulomb gauge condition
manages to remove the gauge freedom in the constraint surface only locally. The axial gauge in general Yang-Mills theory,
$n \cdot A^a = 0$, with $n$ some spacelike four-vector, is another example of a realizable, physical gauge (no Faddeev-Popov
ghosts appear in this gauge), in which manifest relativistic (and rotational) invariance is lost - although the Gribov
ambiguity is actually absent in this gauge.}.\\
Difficulties such as these led \citeN{Dirac} to the idea that, instead of imposing the constraints and gauge fixing conditions
before quantization, a more fruitful approach to the quantization of gauge field theories might consist in performing a
quantization procedure first - i.e. to the unconstrained phase space - and to deal with the constraints and gauge arbitrariness
only at the quantum level. Thus, the algebra of the first class constraints, $\{ \gamma_a , \gamma_b \} = C_{abc} \gamma_c$, 
$1 \leq a, b \leq K'$, for some set of (possibly phase space dependent) real-valued coefficients $C_{abc}$, is replaced by a set of quantum constraints\footnote{Here
it is assumed that the relations (\ref{qconstraint}) make sense. In particular, it is assumed that no operator ordering
ambiguities appear upon quantization of the classical constraint algebra. It is also assumed that the gauge
symmetry is not broken at the quantum level - i.e. no gauge anomalies are present - so that no extra, non-constraint terms
appear on the right-hand side of (\ref{qconstraint}) (or in the commutator of any quantum constraint with the Hamiltonian).
See \citeANP{HenTei}, \emph{ibid}, for further discussion.}
\begin{equation}\label{qconstraint}
[ \hat{\gamma}_a , \hat{\gamma}_b ] \; = \; i \hat{C}_{abc} \hat{\gamma}_c
\end{equation}
and the physical states form a subspace, $\Hilbert_{\mbox{\scriptsize phys}}$, of Hilbert space, which is annihilated by
the constraints
\begin{equation}\label{qconstraint2}
\hat{\gamma}_a | \psi \rangle \; = \; 0 \qquad \quad | \psi \rangle \in \Hilbert_{\mbox{\scriptsize phys}} \quad a = 1, \cdots , K'
\end{equation}
Because of the fundamental linearity of the standard quantum formalism, this way of proceeding - via taking care of the constraints
only at the quantum level, (\ref{qconstraint2}) - might prove to be more tractable.
However, contrary to common belief, for the usual gauge theories of the Maxwell and Yang-Mills type, some of the 
undesirable features that generally appear for physical gauges if these theories are quantized according to the first method 
- commonly referred to as ``reduced phase space quantization'' - actually also arise under generic conditions upon quantizing
these theories according to the second method - commonly referred to as ``Dirac (or covariant) quantization'' -
\emph{even if} non-physical sectors are added to Hilbert space\footnote{Cf. \citeANP{Strocchi} \citeyear{Strocchi2,Strocchi3,Strocchi4}, 
\shortciteANP{FePiSt1} \citeyear{FePiSt1,FePiSt2}, \citeN{StrWig}.}. So, for these theories, there are available two standard quantization
procedures, both of which suffer from certain undesirable features, such as nonlocality or noncovariance, but
of which one seems to have clear \emph{mathematical} advantages. If it could be shown that (under appropriate conditions)
the two procedures would always result in physically equivalent theories, it would be clear which of the two to use.
Unfortunately however, this is not the case and the two quantization procedures in general lead to inequivalent theories\footnote{Cf.
\citeN{AshHor}, \citeN{Gotay2}, \citeANP{Kuchar1} \citeyear{Kuchar2,Kuchar1,Kuchar3}, \citeN{Kunstatter}, \citeN{EppKun}, \citeN{Klauder}.}.
Furthermore, there is no general consensus on which of the two procedures is to be preferred for \emph{physical} reasons in the 
case of actual inequivalence\footnote{In addition to the references cited in the previous footnote, see e.g. also \citeN{Dayi}, \citeN{Earman2}.}
(from a foundational perspective, nontrivial issues are also raised by the presence of gauge symmetries in quantum theory in
general; for instance, the fact that it is necessary to interpret quantum fields as operator valued distributions, strictly
speaking, prohibits any straightforward extension of local symmetry transformations, such as (\ref{gaugetrans1a}) or (\ref{gaugetrans2a}),
to a quantum theory of fields\footnote{For an attempt to rigorously formulate the gauge principle in quantum field theory, see e.g. \citeN{HaagOjima}.}).\\
With regards to the overall argument, the relevance of the preceding remarks on the
quantization of gauge theories is as follows. In the reduced phase space quantization scheme, all information about the gauge
structure of a theory is removed at the classical level, prior to quantization. But the arguments in sections \ref{SSBlocalU1}
and \ref{extension} demonstrated that the reduced phase spaces for the Higgs models considered have different (infinite)
dimensions before and after the phase transition\footnote{On using Wheeler's \citeyear{Wheeler} notation for representing the number of degrees
of freedom for infinite dimensional systems, the Abelian Higgs model for instance is specified by $\infty^{6 \infty^3}$ degrees of
freedom for $m^2 > 0$ and by $\infty^{8 \infty^3}$ degrees of freedom for $m^2 < 0$ (corresponding to respectively three and four
physical fields and their time derivatives, considered at each point of a spacelike hypersurface). Similar remarks apply
to non-Abelian Higgs models.}. Upon a reduced phase space quantization approach, \emph{the conceptual inconsistency 
demonstrated for standard, classical Higgs models thus carries over rigorously to the quantum level}.
On the other hand, if physical priority is assigned to the Dirac quantization method, the arguments of sections \ref{SSBlocalU1}
and \ref{extension} imply that in order for Higgs models to be consistent, direct \emph{physical} information would have to 
be contained in the gauge structure of such models, in a way such that the numbers of physical \emph{quantum} degrees of 
freedom are the same before and after the phase transition. It should be emphasized however, that such a view of things 
amounts to a highly non-standard interpretation of gauge symmetries. Moreover, it turns out that upon adopting the Dirac quantization 
scheme, masses can be generated for gauge fields without the presence of scalar fields\footnote{Cf. \shortciteN{BuEnFe}.}.
Hence, if physical information is contained in gauge structure in the form of \emph{fields}, it would seem that this would
have to be in the form of gauge fields. In treatments based on the covariant method, indirect physical significance is sometimes attributed to quantum gauge degrees
of freedom, but not in the sense that these degrees of freedom are to be regarded as truely physical\footnote{According to
\citeN{PeSch}, for instance, classical gauge artefacts, such as Goldstone bosons, can ``still leave their footprints
in physical observables'', such as decay widths and cross sections (p. 743).\label{gaugefootprints}}. Nevertheless, it is
imaginable that the argument fails upon a - yet to be established - proper assignment of physical degrees of freedom in
covariantly quantized Higgs models. However, as will be seen near the end of this section, in this case there is still 
another nontrivial issue to be addressed.\\
It should also be emphasized here, that in modern treatments based upon Dirac quantization, this method is actually understood
in a generalized sense. That is, according to modern covariant methods, phase space is initially to be \emph{extended} (rather
than reduced or left unmodified), to include even further non-physical degrees of freedom - the Faddeev-Popov ghosts.
However, upon again adopting a standard view of things, this method of BRST quantization is not expected to significantly
affect our main argument, as it should enter the treatment of Higgs models in a symmetric fashion with respect to any phase
transition (in fact, it is not very difficult to set up an explicit argument to this effect within the context of perturbation 
theory). 
All this shows that the conceptual inconsistency of standard, classical Higgs models very likely persists in quantum field
theory. Given the success of a very \emph{specific} Higgs model - the Glashow-Salam-Weinberg (GSW) model - in predicting the weak
intermediate vector bosons, $W^{\pm}$, $Z^0$, first detected in 1983 at CERN, it would however be natural to expect that
this inconsistency of conventional Higgs models reflects the fundamental limitations of these models, rather than a fatal
flaw in the idea that gauge bosons can become massive as a result of the occurrence of a phase transition.\\
Concerning the quantization of gauge theories in general however, it might still be wondered whether some alternative
quantization scheme could not significantly alter the entire argument. In this article, the method of path integral quantization 
is taken to be ultimately based upon the canonical formalism.
However, even if it is assumed that the path integral formalism is somehow more elementary than the canonical formalism
(as seems to be customary in many modern treatments)\footnote{For systems with finitely many degrees of freedom, Feynman's
path integral formalism is (formally) equivalent to ordinary Schr\"odinger quantum mechanics. For systems with infinitely
many degrees of freedom, equivalence with the canonical formalism has not been established, nor is it clear that the two
formalisms \emph{should} in fact be equivalent. One problem is that for some theories, such as the nonlinear $\sigma$-model 
and gauge theories, naive extraction of the Feynman rules directly from the Lagrangian density leads to the wrong results. 
In gauge theories, reference to the canonical formalism is usually sidestepped by employing the Faddeev-Popov method, but
even in that case, the canonical formalism still seems to be necessary for proving unitarity of the $S$-matrix.
See e.g. \citeN{Weinberg2} section 9.1, for further discussion. See also \citeN{Haag}, chapter VIII, or \citeN{Nambu4}.}, 
this is not something that is likely to change the main argument, for reasons already expressed near the end of section \ref{SSBlocalU1}.
It is worth remarking here that the above perspective on the quantization of gauge theories is in basic accordance with
standard treatments, although particular details of representation may differ\footnote{Cf. \citeANP{HenTei}, \emph{ibid}.
In a recent discussion of the gauge principle in Yang-Mills theories, \citeN{Guay} distinguishes four different possibilities
to deal with the gauge surplus structure in the quantization process : (i) fix the gauge and then quantize, (ii) specify the reduced 
phase space and then quantize, (iii) Dirac quantization, and (iv) BRST quantization. He then dismisses possibility (i)
because of the Gribov ambiguity and points out that there is no functional quantization analogue of possibility (iii),
effectively leaving five distinct ``quantization paths'' for Yang-Mills theories (i.e. upon adopting either a canonical
or functional quantization approach). Now, \emph{given} a reduced phase space formulation of a Yang-Mills theory, it is 
true that one can adopt either quantization approach. However, it is usually technically impossible to formulate a given
gauge theory in terms of a reduced phase space, because of difficulties such as the problem to find complete sets of 
observables, or the Gribov obstruction (cf. \citeANP{HenTei}, \emph{ibid}, in particular subsection 2.2.3, appendix 2.A
and subsection 13.2.2). It should also be noted that if Guay's conjecture on the equivalence of
possibilities (ii) and (iv) within a functional quantization approach is correct, this only strengthens our overall argument.}.
Regarding still other quantization schemes (such as e.g. the method of ``deformation quantization''), as far as such schemes
can actually be viewed as constituting viable alternatives to standard quantization, it is 
difficult to see how they could significantly alter the main argument and as already proclaimed, it seems in fact 
much more natural to attribute the discrepancy in physical degrees of freedom directly to a basic flaw in standard \emph{formulations}
of the Higgs mechanism itself. From this perspective, it might appear that a reformulation of the conventional Higgs models in 
terms of a different set of dynamical variables might offer some clues on how to modify these models into an overall consistent 
theoretical framework, in such a way that the main empirical predictions of the conventional models remain unaltered.
Before considering this possibility in some more detail in the next section, it is necessary to address a number of further
issues however.\\
First, as already recalled in section \ref{SSBlocalU1}, according to usual ideas in quantum field theory, a complex scalar field, 
$\phi$, \emph{considered just by itself}, represents two physical states of opposite charge. That is, straightforward application of canonical methods leads to a decomposition of the quantum 
field $\phi$ in terms of positive and negative frequency solutions, with operator coefficients that respectively 
\emph{annihilate} quanta of charge $q$, say, and \emph{create} quanta of charge $-q$; and vice versa for $\phi^{\dagger}$.
So, the quantum theory of just a single, complex scalar field is conventionally taken to describe two types of quanta with opposite 
charge - usually referred to as particles and antiparticles\footnote{See e.g. \citeANP{ItZu}, \emph{ibid}, pp. 120-122, 282-287 for further discussion.}.
The problem with extending this argument to the case of a complex scalar field embedded within the Abelian Higgs model (or, for
that matter, within any theory with the appropriate gauge symmetry), is simply that in that case the scalar field can no longer
be considered just by itself. Classically, it was recalled in sections \ref{SSBlocalU1} and \ref{extension}, that the standard 
view on gauge symmetries dictates that field configurations lying in the same gauge orbit should be physically identified. So, viewed as classical
fields, $\phi$ and $\phi^{\dagger}$, being gauge equivalent to each other, are just different representations of the same
physical field configuration, and classically, there thus appears to be no hope that two physical degrees of freedom can be assigned
to $\phi$ and $\phi^{\dagger}$ in the presence of gauge interactions\footnote{In general, when attempting to determine independent degrees of
freedom for models involving fields, it is important to carefully examine each model in turn, rather than simply take the case of a single complex
scalar field as being representative. For instance, a Dirac spinor, $\psi$, being a complex-valued 4-component object, only
carries four physical degrees of freedom - corresponding to two types of quanta of opposite charge, upon quantization, which 
can have half a unit of spin angular momentum either up or down in some arbitrary reference direction - rather than the eight 
states expected from the analogy with the complex scalar field. This is because the first order time derivative,
$\dot{\psi}$, or, alternatively, the canonical momentum conjugate to $\psi$, are not independent of $\psi$.}.
If the field $\phi$ and its complex conjugate are interpreted as (distributional) quantum fields, the situation becomes obscured 
by the nontrivial nature of gauge symmetries in quantum theory - already alluded to previously. 
However, within a path integral approach, as discussed above, a semi-classical analysis is certainly appropriate and there
again seems to be little room to escape the conclusion that $\phi$ and $\phi^{\dagger}$ are in the same gauge class, leading to
only one physical degree of freedom being associated with the complex scalar field. One thus again ends up with the by now familiar 
argument on the physical consistency of Higgs models, but this time obtained within a setting that is arguably quantum field theoretical.\\
Given the importance of the distinction between globally and locally invariant models in connection to the number of independent 
degrees of freedom, as just emphasized, it could be wondered whether the discrepancy for conventional Higgs models
could not simply be resolved within the usual context of perturbative quantum field theory, more or less as follows.
\noindent If the gauge couplings are tuned to zero, the scalar fields decouple from the gauge fields (cf. eqs.
(\ref{covder}), (\ref{ssbL3}), for the case of just one such coupling) and Goldstone bosons re-enter the physical spectrum.
If the physical spectrum at small, nonvanishing gauge couplings is viewed in some way as a ``quantum correction'' to the
leading order spectrum at zero gauge couplings, this spectrum can then be heuristically accounted for in terms of the
standard picture of ``Goldstone bosons having been eaten by the gauge bosons and as a result became heavy''.
Such explanations of the Higgs mechanism sometimes seem to be vaguely hinted at in more sophisticated discussions\footnote{See
e.g. \citeN{Weinberg3}, chapter 21, or \citeANP{tHooft}, \emph{ibid}, section 5.3.}. However, although perhaps useful for heuristic
purposes, these kind of descriptions of the Higgs mechanism cannot constitute a viable picture of what is \emph{actually} going on
at the phase transition, at which the gauge couplings have some definite nonzero values and Goldstone scalars never enter
the physical spectrum in the first place. Moreover, in a consistent perturbative description it would also be necessary
to view effects at small nonzero $\lambda$ as quantum corrections to the situation at zero scalar field coupling, but 
the scalar field minimum diverges at zero perturbative order, as $v^2 \sim \lambda^{-1}$, whereas the vanishing of the
gauge boson masses in the free case would seem to depend on the order in which the limits $g \rightarrow 0$ and $\lambda \rightarrow 0$
are taken. Related to this is the fact that the picture presented in many textbooks, of the quantum scalar field, $\phi$,
picking up a nonzero expectation value, $v$, at the phase transition, i.e. $\langle \phi \rangle = v$, contradicts the
basic fact of standard perturbation theory that all odd $n$-point functions for $\phi$ necessarily vanish.
From the perspective of the scalar self-coupling, it thus appears that the Higgs phenomenon is intrinsically nonperturbative\footnote{See
also \citeN{AitHey}, section 13.3. One could of course simply add a constant $v$ to a quantized field $\phi$ to obtain
an expectation value $v$ for $\phi$, but it is not clear what this means : the resulting quantity is neither a strong solution
to the equation of motion, nor a normal-ordered, weak one. It should also be mentioned here that even at the nonperturbative
level, the statement that $\langle \phi \rangle$ is nonvanishing cannot be given a gauge independent meaning, strictly speaking, 
as a consequence of Elitzur's theorem. See also section \ref{altview}.}.
A somewhat related issue is the following. According to the so-called Goldstone boson equivalence theorem, at high energies
the amplitude, $T(W^+_L,W^-_L,Z^0_L,H)$, for scattering of longitudinal intermediate bosons, $W^{\pm}_L$, $Z^0_L$ and physical
Higgs bosons, $H$, in the GSW model, effectively becomes equal to the amplitude, $T(w^+,w^-,z^0,h)$,
for scattering of the massless Goldstone bosons $w^{\pm}$, $z^0$ and neutral massive boson, $h$, that are part of the physical
spectrum of the model at zero gauge couplings (cf. the preceding remarks). This theorem is a crucial ingredient in purported
proofs that the only renormalizable quantum field theories with massive gauge bosons are essentially the theories with
spontaneously broken gauge symmetries\footnote{\shortciteN{CoLeTi}, \shortciteN{LeeQuiTha}, \citeN{Llewellyn}.
See also \citeANP{PeSch}, \emph{ibid}, section 21.2.}.
But what is the meaning of the theorem itself ? It appears that the underlying picture implicitly adopted here is that
at high energies the gauge couplings are tuned to zero. But according to the usual picture, at high energies the theory
is in the \emph{unbroken} symmetry phase without any vacuum degeneracy and accompanying Goldstone bosons whatsoever.
Moreoever, for the standard model, the gauge couplings $g$ and $g'$ (corresponding to respectively $\mbox{SU}(2)$ weak isospin
and weak hypercharge) are related to the unit of ordinary electric charge, $e$, according to $e = g \sin \theta = g' \cos \theta$,
where $\theta$ denotes the Weinberg angle. Yet, the fine structure constant, $\alpha \simeq e^2 / 4 \pi$, which is approximately
equal to $1/137$ at the atomic scale, is known to increase with increasing energy, both experimentally and theoretically (in
fact, there is a well known - although not entirely conclusive - argument according to which $\alpha$ diverges at some
ultrahigh energy scale). So, there is certainly no good reason to suppose that \emph{both} gauge couplings $g$ and $g'$ tend 
to zero in the far ultraviolet region.\\
As already mentioned before, it is imaginable that a - yet to be established - proper assignment of physical degrees of freedom 
in covariantly quantized Higgs models can be given. Quite generically, it is necessary to acknowledge the existence of a 
``subphysical level'' of representation associated with such models, in which entities that are usually taken to be non-physical, such as the 
Goldstone scalars and Faddeev-Popov ghosts, also play a crucial role at the intermediate stages in calculations of amplitudes
for physical processes. The point is now that in typical modern references, these entities - although still regarded as ``non-physical'' 
- are also envisaged to have nontrivial effects on physical observables of the theory\footnote{An example of this was already given in note \ref{gaugefootprints}.
As another example, \citeN{Jackiw} has contemplated the existence of a ``ghost effect''.}.
In fact, this is not so very surprising. If the effects of these entities on physical observables were always completely
trivial, it would seem difficult to avoid the conclusion that the role of gauge symmetry is more formal and merely serves 
to select the allowed terms in the action. In particular, this would then seem to indicate an equivalence with the reduced
quantization method, but it was already pointed out that no such equivalence in general exists. Moreover, as also seen
already, within the reduced quantization scheme the classical argument for inconsistency of the usual Higgs models carries over
rigorously to the quantum case, while the whole point of the present discussion is to consider a quantized Higgs model
for which this argument actually does not carry over.\\
The situation here may be characterized in terms of certain elements of the ``surplus structure'' of quantized gauge theories
gradually being shifted into the physical structure of these theories\footnote{As emphasized by \citeN{Redhead}, the sharp
boundary between the representation of the ``physical structure'' and the ``surplus structure'' within the total mathematical
structure used to represent a given theory, may get blurred over time (cf. the notion of energy in nineteenth-century physics).}.
However, in order that this picture be a completely consistent one, it appears to be necessary to also allow for an arbitrary
ground state at the subphysical level of representation. Indeed, without this condition there could be no Goldstone modes
at this level either. Thus, again taking the Abelian case to be representative, a generic perturbed ground state would have to be
of the form (\ref{perturb1}), with $\alpha \in \real{}$. Now, the constant phase $\alpha$ does not have any effect
in (\ref{ssbL2}), but it \emph{does} appear in the Yukawa-interaction term $\sim \lambda_f (\bar{\psi}_L \phi \psi_R + \bar{\psi}_R \phi^{\ast} \psi_L)$,
for the scalar field and the chiral fermions, $\psi_L$, $\psi_R$. The point is now that if it is argued
that unphysical particles can leave their footprints in physical observables, there seems to be no reason whatsoever to suppose
that the same is not true for the generic ground state phase $\alpha$. However, in the analogous case of the GSW-model,
for which it is envisaged that the underlying electroweak symmetry of physical laws was in actuality spontaneously broken
in the very early universe, this would mean that causally disconnected spacetime regions would all have had to ``choose''
exactly the same value for the ground state phases, in order not to lead to inconsistencies with observation at the \emph{present} time\footnote{See \citeN{Penrose}, section 28.3, for a more detailed discussion of this point.}.
Within such a view of things - apart from the fact that it would still be necessary to properly identify the correct physical
degrees of freedom in this case - the EPR type of nonlocal behaviour, so familiar from ordinary, composite quantum spin systems, would
then also play a crucial role in making such a view consistent in the first place. Quite interestingly, upon adopting
a more conservative, i.e. standard, reading of gauge symmetries, nonlocality also appears to be unavoidable in rendering Higgs
models conceptually consistent. This is discussed next.

\section{Further Possible Ways of Resolving the Discrepancy}\label{altview}

\noindent As already mentioned several times, the impossibility of correctly matching the degrees of freedom for the two
phases of Higgs models, arises for standard formulations of such models and upon adopting the usual interpretation of
gauge symmetries. The difficulty may therefore disappear upon relaxing at least one of these two conditions.
A general approach for realizing the first option, i.e. to obtain a non-standard account of the Higgs mechanism, goes
back to a gauge invariant formulation of quantum electrodynamics due to \citeN{Dirac3}.
The basic idea in this formulation is to redefine the classical field variables, so that they become gauge invariant.
For instance, in the case of a complex scalar field, $\phi$, coupled to an Abelian gauge potential, $A_{\mu}$, as prescribed
by (\ref{ssbL2}), one can introduce the variable $\Phi$, defined by
\begin{equation}\label{Diracvariable}
\Phi (x) \; = \; \exp \left( - i g \int \! d^4 x' \, \chi^{\mu} (x , x') A_{\mu}(x') \right) \phi(x)
\end{equation}
for some real vector function $\chi^{\mu}$.
Upon performing a simultaneous gauge transformation, (\ref{gaugetrans1a}), (\ref{gaugetrans1b}), one finds that $\Phi$
transforms according to
\begin{equation}
\Phi(x) \longrightarrow \Phi' (x) \; = \; \Phi(x) \exp \left( i \alpha (x) \: + \: i \int \!  d^4 x' \, \alpha (x') \partial_{\mu}' \chi^{\mu} (x , x') \right)
\end{equation}
and hence $\Phi$ is gauge invariant for generic $\alpha$ if and only if 
\begin{equation}\label{gaugecond}
\partial_{\mu}' \chi^{\mu} (x , x') \; = \; - \delta^4 (x - x')
\end{equation}
(provided also that the gauge functions are of compact support). Assuming that these conditions are met, there is now no
difficulty that the variable $\Phi$ and its complex conjugate, $\Phi^{\dagger}$, lie in the same gauge orbit, as these variables
are gauge invariant. Hence, there is a nonambiguous way in which these variables can be taken to represent independent degrees 
of freedom, which moreover precisely reduce to those of the globally invariant case on putting $A=0$. 
Let us now see how the foregoing remarks might in principle be of use in obtaining a solution to the problem of the 
degrees of freedom in Higgs models. Following this, the discussion will then turn to some of the obstacles encountered by
this general approach.\\
Continuing with the Abelian case, an immediate observation to be made is that that the potential (\ref{ssbV1}) retains the exact same form in
terms of the new variables, i.e. $V(|\Phi|^2) = V(|\phi|^2)$. But since the new variables are gauge invariant, there is now
no difficulty that the vacuum degeneracy of the potential for $m^2 < 0$ disappears upon gauging the $\mbox{U}(1)$-symmetry,
as was seen to be the case for the original field variables, $\phi$ and $\phi^{\dagger}$.
That is, although the new variables are gauge invariant, they are not invariant under \emph{global} $\mbox{U}(1)$-rotations
and as far as the scalar field potential is concerned, the net effect of the transformation of field variables is thus to
factor out the gauge freedom and to leave a residual global $\mbox{U}(1)$-symmetry - which can be spontaneously broken.
The situation is thus effectively reduced to that of section \ref{SSBglobalU1}, but this time also with gauge fields present.
Skipping specific details\footnote{Cf. \citeN{Lusanna}, \citeANP{LuVa1} \citeyear{LuVa1,LuVa2}. For a more generic description
of what is involved, see \citeANP{Earman1} \citeyear{Earman1,Earman3}.}, the idea is now to transform to yet another set
of variables in the broken symmetry phase, $m^2 < 0$, so that the fields $\Phi$, $\Phi^{\dagger}$ mix with the gauge field
variables, with the result of yielding one Higgs mode and one Goldstone mode, with the latter subsequently identified with
the longitudinal polarization component of the massive gauge field.
Assuming that all specific details can be consistently filled in, also in the non-Abelian case, the above gauge invariant
formulation would thus seem to completely resolve the conceptual consistency problem of classical Higgs models.
Although this may indeed well be the case, one thing that should be noted is that it is not at all clear whether this formulation
corresponds to ``actual physics'' in any way - something which would certainly be required in order to count as a complete
resolution of the consistency problem (recall also note \ref{reshuffle} in this regard).
One apparent drawback, from the viewpoint of classical field theory, is the intrinsic nonlocal character of the scalar field
variables, as evident from the definition (\ref{Diracvariable})  - more on this shortly.\\
There are many solutions to eq. (\ref{gaugecond}). Taking $\chi^{\mu}(x,x') = \delta (x^0 - x'^0) c^{\mu}(\vect{x}{},\vect{x}{}')$, with
$c^0 (x,x') := 0$, corresponds to the case originally considered by Dirac, for which eq. (\ref{gaugecond}) reduces to
\begin{equation}\label{gaugecond1}
\partial_{i}' c^i (\vect{x}{} , \vect{x}{}') \; = \; - \delta^3 (\vect{x}{} - \vect{x}{}')
\end{equation}
and which still has many solutions - one being the familiar gradient form (considered in detail by Dirac),
$c^i (\vect{x}{} , \vect{x}{}') = \vect{x}{} - \vect{x}{}' / (4 \pi |\vect{x}{} - \vect{x}{}'|^3) = \vectr{\nabla}{}' 1 / (4 \pi |\vect{x}{} - \vect{x}{}'|)$.
The gauge invariant variable may then be formally expressed as (with $1/\nabla^{2}$ denoting the inverse of the three-dimensional Laplacian)
\begin{equation}\label{Diracvariable1}
\Phi (x) \; = \; \exp \left( - ig \vectr{\nabla}{} \cdot \vect{A}{} / \nabla^2 \right) \phi (x)
\end{equation}
which is manifestly noncovariant.
Another possibility is to take
\begin{equation}\label{Mandelstam1}
\chi^{\mu} (x , x') \; = \; \int_{0}^{1} \! ds \, \delta^4 (x' - \xi (s)) x'^{\mu}
\end{equation}
where $\xi^{\mu} : [0,1] \rightarrow \real{4}$ denotes any smooth path in spacetime from spatial infinity to $x$.
Although it is not immediately obvious that this choice actually defines a solution to eq. (\ref{gaugecond}), it is readily verified
that (\ref{Mandelstam1}) indeed yields a gauge invariant variable when substituted into eq. (\ref{Diracvariable})\footnote{See 
also \citeN{Dirac3}, p. 659, for the consideration of a special case of this scenario. The variable (\ref{Mandelstam}) appears to have 
been introduced independently by \citeN{Mandelstam} and the spacetime path appearing in the exponential is now usually referred 
to as a \emph{Mandelstam string}.}
\begin{equation}\label{Mandelstam}
\Phi (x) \; := \; \exp \left( - ig \int_{-\infty}^x \! d \xi \cdot A (\xi) \right) \phi (x)
\end{equation}
In view of the proliferation of solutions to eq. (\ref{gaugecond}), a difficulty for obtaining a resolution to the Higgs 
consistency problem is thus to somehow extract the correct solution, or a set of such solutions that are all physically
equivalent in some way. But upon temporarily ignoring this point, suppose that the accompanying variables can be quantized 
in the usual way - as is true for the cases originally considered by Dirac and by Mandelstam. The intrinsic nonlocality of 
the scalar field variable (and its generalizations) would then not seem such a serious problem, since it was argued in the 
previous section that the quantization of gauge theories by either of the two main schemes is in general accompanied by 
nonlocal and/or noncovariant features. However, the nonlocal character of Dirac variables in general causes the Lagrangian
densities formulated in terms of these variables to become nonlocal (and also nonpolynomial), as a result of which standard
renormalization and regularization prescriptions no longer apply\footnote{See \citeN{LuVa1} for some remarks on how this difficulty
might be overcome. It should be recalled that one of the main reasons why the GSW model gained acceptance by physicists in
the early 1970s was precisely because it turned out to be a renormalizable model.}. Moreover, there are also still some other difficulties.\\
Within the context of lattice gauge field theory - which is formulated entirely in gauge invariant terms and which forms the only known
constructive approach towards a nonperturbative treatment of gauge theories in general\footnote{See e.g. \shortciteN{FreReSe}
for a brief review of the lattice approach to gauge theories.} - Elitzur's theorem states that all nonvanishing correlation functions must be gauge invariant\footnote{Cf. \citeN{Elitzur}, \shortciteN{DDG}.}.
In particular, the vacuum expectation value of any gauge dependent quantity, such as the original scalar field variable, $\phi$,
in the above treatment, is necessarily zero and spontaneous breaking of local symmetries can therefore not occur, according to this theorem.
Within the usual context of perturbative formulations of gauge field theories, a gauge fixing scheme is typically introduced
and Elitzur's theorem no longer applies. But, although $\langle \phi \rangle$ is then in general nonvanishing, it would
be inappropriate in this case to speak of \emph{spontaneous} symmetry breaking, in general, since the symmetry breaking is explicit,
because of the gauge fixing. Nevertheless, some gauge fixing schemes may still leave a residual \emph{global} symmetry and
it can then be asked whether this remnant symmetry can be spontaneously broken in the usual sense.
As it turns out, the answer to this question closely depends on the specific details of the gauge fixing procedure and within 
this context, the spontaneous symmetry breaking is therefore nothing but a gauge artefact\footnote{For the Abelian Higgs model, 
spontaneous breaking of a remnant symmetry turns out to be possible for the Landau gauge, for instance, but not for other gauges. See \citeN{KennedyKing}.}.\\
These conclusions are in agreement with those obtained by an alternative approach, based on the Dirac observables (\ref{Diracvariable}),
in which the choice of gauge has been built in via the solutions to eq. (\ref{gaugecond}). For instance, the case originally
considered by Dirac corresponds to the Coulomb gauge, in the sense that the vacuum expectation value of the locally (but not globally)
$\mbox{U}(1)$-invariant variable $\Phi$, as given by (\ref{Diracvariable1}), equals the vacuum expectation value of the original
field variable $\phi$, with the Coulomb gauge imposed. Other solutions $\chi^{\mu}$ correspond to other gauge fixing schemes,
but different gauge choices lead to different remnant symmetries, which are in general spontaneously broken at different
loci in coupling constant space\footnote{Cf. \citeN{CaudyGreen}.}.\\
More generally, \shortciteANP{FroMoStro} \citeyear{FroMoStro,FroMoStro2} have argued that the (gauge fixing scheme dependent)
nonvanishing character of order parameters (such as $\langle \phi \rangle$) in standard, perturbative formulations is due to the omission of nonperturbative
effects and that in particular, upon including topological defects, such as instantons, vortices, etc., order parameters
generically vanish. Furthermore, according to this treatment it is possible to set up a general, gauge invariant formulation
which explains the empirical success of the GSW model in its standard perturbative formulation,
via the recovery (to a certain extent) of gauge dependent correlation functions as calculated in standard perturbation expansions,
from the corresponding gauge independent correlation functions of certain gauge invariant variables (which are different from 
those in the Dirac approach). With respect to our overall argument however, the apparent necessity to include solitonlike effects in order\enlargethispage*{5cm}
to obtain a conceptually consistent picture of the Higgs mechanism within this kind of setting, must certainly be regarded as a non-standard procedure.
Moreover, the appearance of such effects in the argument raises further interpretational issues in itself. For instance, for
gauge field theories it is unclear whether instantons can be interpreted realistically, since the only allowed instanton solu-
\newpage
\noindent tions in Minkowski spacetime for such theories are associated with noncompact groups, such as $\mbox{SL}(n,\complex{})$.
Moreover, the frequently encountered viewpoint that gauge instantons solve the so-called $\mbox{U}(1)$-problem in QCD
is not conclusive, since there is an alternative formulation of QCD (or of gauge theories in general, for that matter),
which solves this problem without invoking any instantons \emph{and} without giving rise to the ``strong CP problem'' - 
unlike the standard formulation\footnote{Cf. \citeN{FortGambini}. In addition to an approximate global chiral
$\mbox{SU}(2) \times \mbox{SU}(2)$ symmetry (cf. note \ref{SSBQCD}), strong nuclear interactions also display an approximate global chiral
$\mbox{U}(1)$ symmetry, which is not displayed in physical states. If this symmetry were spontaneously broken, there would have to be an additional near-Goldstone boson
associated with the strong interactions, with a mass comparable to that of the pions - but no such boson is known to exist.
The only viable candidate, the $\eta$ meson, is considerably heavier than the pions. Instantons (viewed as tunneling events 
between discrete, degenerate QCD-vacua) could generate a substantially larger mass for the $\eta$, but they would also give
rise to an additional parameter, the QCD-vacuum angle, $\theta$. The fact that, experimentally, this angle is known to be 
extremely small, while being unconstrained theoretically, is known as the strong CP problem.}.

\section{Discussion}\label{discussion}

\begin{quote}
There is much more danger in a breakdown in communication among scientists than in a premature consensus that happens to
be in error. It is only when scientists share a consensus that they can focus on the experiments and the calculations that
can tell them whether their theories are right or wrong, and, if wrong, can show the way to a new consensus.\\
Steven Weinberg \citeyear{Weinberg6}
\end{quote}

\noindent Summing up, although there is certainly room to evade the main conclusion concerning physical degrees of freedom in Higgs 
models of sections \ref{SSBlocalU1} and \ref{extension} by considering non-standard formulations of such models (or non-standard
interpretations of gauge symmetry), it was also seen in the previous two sections that such non-standard formulations (or
interpretations) give rise to additional difficulties.
There thus remains a serious challenge to be met as far as attaining a proper understanding of the Higgs mechanism is concerned. 
The present work should be seen in this light : its value is that it clearly exposes a conceptual flaw in conventional formulations
of the Higgs mechanism, thereby rendering the problem of understanding the exact meaning of this mechanism a very pressing 
problem indeed. It is precisely because of this point that the present results are considerably stronger than the ramifications
of two other thorough, critical reviews of the Higgs mechanism that saw the light in recent years.\\
In particular, \citeN{Smeenk} focusses on the two issues of extracting the gauge invariant content of the Higgs mechanism 
and the (lack of) meaning of the notion of a ``spontaneously broken local symmetry'' and concludes that a gauge invariant
description of the Higgs mechanism is not impossible, as a matter of principle. \citeANP{Lyre}'s \citeyear{Lyre} perspective
is slightly different. He focusses on standard, perturbutive formulations of the Higgs mechanism and concludes that such 
formulations cannot be representative of a genuine physical process, as they merely reshuffle degrees of freedom.
Although much in these two reviews is in general agreement with what is said here, both reviews unfortunately fail to
recognize the main point of the present work. Smeenk does not actually address the issue which leads to this point, but this 
does not cause his conclusions to contradict those presented here. With regards to Lyre's work, it should be clear that his conclusion based on an actual reshuffling of degrees of
freedom disagrees with the main claim of the present work. However, even is this were not the case, it must be asked whether the argument that a 
``mere'' reshuffling of degrees of freedom cannot be representative of a genuine physical process is in itself conclusive.
This does not appear to be the case (recall in particular also note \ref{reshuffle} in this regard).
In addition, since Lyre's treatment is entirely classical, it is unable to take into account the possibility of degrees of
freedom in the surplus structure (\`a la Goldstone or Faddeev-Popov-BRST) acquiring some sort of physical status at the 
quantum level (recall the discussion in section~\ref{qftcontext}).\\
It is also worth noting that non-standard formulations of the Higgs mechanism have been known to exist already for quite some 
time. For instance, Ferrari and co-workers have considered specific models exhibiting spontaneously broken gauge symmetries with 
(i) massive gauge fields, but without a Higgs field\footnote{Cf. \shortciteN{BuEnFe}.}, and (ii) (massless) scalar fields, 
but without massive gauge fields\footnote{Cf. \citeN{Ferrari}.}, something which strongly suggests that different mechanisms 
are actually operative in symmetry breaking and in generating masses for the gauge bosons\footnote{See also \shortciteN{BuEnFe2}.}.
This latter feature is consistent with the results of \citeN{StollThies}, who consider an approach towards the Higgs phenomenon 
in which redundant (i.e. gauge) degrees of freedom are eliminated from the start and in which the residual (global) symmetry is unitarily 
implemented in the Higgs phase, while photons are interpreted as Goldstone bosons of the spontaneously broken residual symmetry 
in the ``Coulomb phase''.\\
Finally, it is interesting to make a comparison with the phenomenon of superconductivity, as it is often stressed - particularly
in standard, perturbative formulations - that, as far as basic underlying physical principles are concerned, there is no
difference between the Higgs mechanism and a superconducting order transition. From our perspective however, it is not
difficult to check that the analogy with superconductivity is unable to shed much light on the problem of the unaccounted 
degrees of freedom - quite to the contrary in fact. Nevertheless, it does suggest quite strikingly
that there may not be a fundamental Higgs field at all, but merely an effective Landau-Ginzburg type of field, which is entirely
absent at high energies (i.e. energies above the Fermi scale) and which is perhaps some bound state of (new species of) fermions, as in technicolour models\footnote{See
e.g. \citeN{Mohapatra}. As is well known, renormalization group arguments strongly indicate that the quantum field theory
defined by (\ref{ssbL1})-(\ref{ssbV1}) is \emph{trivial} (but see \citeN{Callaway}). This is usually interpreted to mean
that the Higgs field is really only an effective field, which loses its meaning at some high energy scale $\Lambda$ (quite
analogous to the Landau-Ginzburg order parameter). If - as is usually assumed to be the case - the Higgs is weakly coupled at
the electroweak scale, data from electroweak precision measurements strongly favour a relatively light Higgs, and $\Lambda$
turns out to be of Planckian orders. However, this gives rise to enormous fine-tuning problems in calculations, for which
supersymmetry is typically called upon as a solution. If the Higgs field is not weakly coupled, its mass could be larger and
$\Lambda$ correspondingly smaller. Apart from \emph{very specific} theoretical desiderata, there does not seem to be any
reason to suppose that the Higgs field preserves its identity all the way up the Planck scale - which is after all still
some seventeen orders of magnitude away from scales currently experimentally accessible. As stressed by \citeN{Veltman2},
the ``Higgs mass prediction from present data may be a joke''.}.

\section*{Acknowledgements}

\noindent The author thanks Damien George, prof. Ruggero Ferrari, prof. Michael Thies and two anonymous referees for 
some helpful comments.

\bibliographystyle{eigen}
\bibliography{higgsv2}

\end{document}